\documentclass[article,aps,prr,twocolumn,superscriptaddress,english,longbibliography]{revtex4-2}

\usepackage[usenames,dvipsnames]{xcolor}
\usepackage[colorlinks=true,citecolor=blue,urlcolor=blue,linkcolor=blue]{hyperref}
\usepackage{etex}
\usepackage{amsmath,amssymb,amsthm}
\usepackage{times,txfonts}
\usepackage{braket}
\usepackage{color}
\usepackage{natbib}
\usepackage{amsmath}
\usepackage{mathtools}
\usepackage{latexsym}
\usepackage{tabularx, booktabs}
\usepackage{graphics,epstopdf}
\usepackage{graphicx}
\usepackage{float}
\usepackage{amsfonts}
\usepackage{tikz}
\usepackage[ruled]{algorithm2e}
\usetikzlibrary{quantikz}
\usepackage{color,soul}

\newcommand{\be}{\begin{equation}}
\newcommand{\ee}{\end{equation}}
\newcommand{\ba}{\begin{eqnarray}}
\newcommand{\ea}{\end{eqnarray}}

\usepackage{multirow}
\usepackage{appendix}
\usepackage{url}
\renewcommand{\arraystretch}{1.7}

\usepackage{orcidlink}
\usepackage{mathtools}
\usepackage{makecell,tabularx}
\setcellgapes{3pt}
\usepackage{scalerel} % For \scaleto

\begin{document}

\title{Spatial Form Factor for Point Patterns: Poisson Point Process, Coulomb Gas, and Vortex Statistics}
\author{Matteo~Massaro\orcidlink{0009-0004-0935-8022}}
\email{matteo.massaro@uni.lu}
\affiliation{Department  of  Physics  and  Materials  Science,  University  of  Luxembourg,  L-1511  Luxembourg,  Luxembourg}

\author{Adolfo~del Campo\orcidlink{0000-0003-2219-2851}}                 
\affiliation{Department  of  Physics  and  Materials  Science,  University  of  Luxembourg,  L-1511  Luxembourg,  Luxembourg}
\affiliation{Donostia International Physics Center,  E-20018 San Sebasti\'an, Spain}

\begin{abstract}
Point processes have broad applications in science and engineering. In physics, their use ranges from quantum chaos to statistical mechanics of many-particle systems.
We introduce a spatial form factor (SFF) for the characterization of spatial patterns associated with point processes. Specifically, the SFF is defined in terms of the averaged even Fourier transform of the distance between any pair of points. We focus on homogeneous Poisson point processes and derive the explicit expression for the SFF in $d$-spatial dimensions. The SFF can then be found in terms of the even Fourier transform of the probability distribution for the distance  between two
independent and uniformly distributed random points on a $d$-dimensional ball, arising in the ball line picking problem.
The relation between the SFF and the set of $n$-order spacing distributions is further established. The SFF is analyzed in detail for $d=1,2,3$ and in the infinite-dimensional case, as well as for the $d$-dimensional Coulomb gas, as an interacting point process. As a physical application, we describe the spontaneous vortex formation during Bose-Einstein condensation in finite time recently studied in ultracold atom experiments and use the SFF to reveal the stochastic geometry of the resulting vortex patterns. In closing, we also introduce a generalization of the SFF applicable to arbitrary sets in a metric space.
\end{abstract}

\maketitle 
    
\section{Introduction}
A point process is a probabilistic model used to describe the random distribution of points in a given space, making it a versatile tool with wide-ranging applications across fields such as physics, engineering, computer science, finance, ecology, and epidemiology \cite{Chiu2013stochastic,Torquato18}. 
By way of example, in physics, point processes play a prominent role in the statistical description of classical many-particle systems such as liquids and gases \cite{Hansen2013,allen2017}.  In this context, the probability distribution of the spatial configurations of the particles 
at thermal equilibrium is governed by a Boltzman factor involving the interparticle potential, realizing an interacting point process.

Among the various types of point processes, the Poisson Point Process (PPP) stands out as the most widely studied due to its simplicity and analytical traceability. Loosely defined, a PPP is a completely random process with no interaction between points. 
In quantum physics, PPPs play a significant role in the analysis of quantum chaos \cite{Guhr1998, Haake2010}. According to the Bohigas-Giannoni-Schmit conjecture, the spectral properties of quantum chaotic systems are governed by random matrix theory \cite{BGS1984}. A key aspect of diagnosing quantum chaos involves the statistical analysis of a Hamiltonian ensemble, focusing on metrics such as eigenvalue spacings and the spectral form factor—essentially, the Fourier transform of the eigenvalue distribution \cite{Leviandier86, WilkieBrumer91, Alhassid93, Ma95}. Depending on the symmetries of the Hamiltonian, an established correspondence exists between spectral statistics and PPPs, further highlighting the significance of PPPs in the study of quantum systems \cite{Haake2010, Sakhr2006}.

Another important application of PPPs, as well as interacting point processes, arises in nonequilibrium statistical mechanics, particularly in the study of phase transitions. It is well known that the occurrence of a phase transition in finite time leads to the formation of topological defects \cite{Manton2004} according to the celebrated Kibble-Zurek mechanism \cite{Kibble76a,Kibble76b,Zurek96a,Zurek96c,delcampo2014}. Defect formation similarly occurs at fast quenches \cite{Chesler2015,Zeng2023,Xia2024}.  
Examples of such defects include kinks in scenarios of parity-symmetry breaking \cite{Vachaspati2023,Laguna98,delcampo10,GomezRuiz20,Suzuki24} and vortices in the case of $U(1)$ symmetry breaking, e.g., during the formation of a Bose-Einstein condensate \cite{Weiler08,Navon2015,Goo2021,Goo2022}, a strongly-coupled superconductor \cite{delCampo2021} or a superfluid \cite{Chesler2015,Xia2024}, such as a unitary superfluid Fermi gas \cite{Ko2019,lee2023observation}. Point processes have been proposed for modeling the spatial patterns of point-like topological defects spontaneously formed across a phase transition \cite{delcampo22,thudiyangal2024}. Specifically, a PPP with intensity set by the Kibble-Zurek mechanism has been shown to accurately describe the statistical features of spatial patterns of kinks and vortices in the early stages after completing the phase transition. As the time of evolution goes by, interactions between defects become increasingly important and alter the spatial statistics of topological defects. The analysis of such patterns motivates our study.

The characterization of point processes is generally done by studying several complementary measures derived from the $n$-point probability distribution, such as pair distribution functions, void probabilities, and contact distributions \cite{Chiu2013stochastic}. Among them,  the structure factor arises naturally in the analysis of scattering patterns. It is defined as the Fourier transform of the pair distribution function.
The structure factor provides insights into the degree of order, randomness, or clustering in the system. As a Fourier transform, it provides a global measure of a complex pattern,  facilitating its study  by identifying salient frequencies, revealing its structure and symmetries. 

In this work, we introduced the Spatial Form factor (SFF) as a complementary measure to the structure factor. The SFF is analogous to the spectral form factor in many-body physics, a diagnostic tool widely studied in quantum chaos and integrable systems.
The SFF is defined as the averaged even Fourier transform of the distance between any pair of points. We motivate this measure in Sec. \ref{SecMotivation}. Section \ref{SecGFFdef} introduces its definition and presents numerical examples for regular, quasi-periodic, and random patterns. It further establishes the dual relations between the SFF and distance distribution. The computation of the SFF for a homogeneous Poisson point process is presented in Sec. \ref{GFFPPP}. Section \ref{SecCharGFF} is devoted to the characterization of the SFF, including its asymptotics, evaluation in special cases, connection to quantum chaos, higher-order spacing distributions, and large-dimensional limit. In Sec. \ref{sec:SFF_Coulomb} the SFF is extended to interacting point processes, using the Coulomb gas as a notable example. In the one-dimensional case, a high-temperature perturbative expansion is carried out, yielding the first-order correction to the SFF due to the Coulomb interaction. Additionally, the SFF is analyzed in higher dimensions using Monte Carlo simulations.
In Sec. \ref{Sec:Vortex_Formation}, we highlight the usefulness of the SFF in characterizing the spatial patterns of vortices that spontaneously form during the rapid cooling of a Bose gas.
We close with a summary and discussions in Sec.  \ref{SecSummary}, in which we introduce a generalization of the SFF to arbitrary sets in a metric space.
%, lifting the restriction to sets of point coordinates. 

\section{Motivation for the Spatial Form Factor}\label{SecMotivation}

In the study of liquids and solids, the structure factor is a useful tool to analyze and identify different phases of matter \cite{allen2017}. It is defined as
\begin{equation}\label{def_S(k)}
S(\mathbf{k}):= \frac{1}{N} \left\langle \sum_{i,j=1}^{N} e^{i\mathbf{k} \cdot (\mathbf{r}_i - \mathbf{r}_j)} \right\rangle,
\end{equation}
where $\mathbf{r}_{i}$ denotes the position of the generic $i$-th constituent of the sample system, such as an atom in a gas or on a crystal lattice. Typically, the positions of the $N$ constituents, which we generically refer to as particles, are not static but change over time. Therefore, the definition of $S(\mathbf{k})$ involves averaging $\langle \cdot \rangle$ over different configurations, determined by the varying particle locations measured within a finite time window, or more generally, drawn from a given ensemble. Statistical mechanics provides a natural framework for a theoretical description of liquids and gases. In this context, the structure factor is often expressed in terms of the Fourier transform of the radial distribution $g(\mathbf{r})$ or the pair correlation function $h(\mathbf{r})=g(\mathbf{r})-1$ as \cite{Hansen2013,allen2017}
\begin{equation}\label{def_S(k)}
S(\mathbf{k}):= 1+\frac{N(N-1)}{N}\int d\mathbf{r} e^{-i\mathbf{k} \cdot \mathbf{r}}g(\mathbf{r}).
\end{equation}

When particle positions follow a periodic pattern, as in a crystal, $S(\mathbf{k})$ exhibits prominent peaks, indicating such regularity.
In contrast, the opposite scenario, typical of an ideal gas, is characterized by a lack of fine structure due to the absence of interactions. In this case, the system can be modeled as a spatial point process, where points are randomly, uniformly, and independently scattered within the system domain for each configuration.
We will primarily focus on two types of such stochastic processes: the binomial point process (BPP) and the homogeneous Poisson point process (PPP), for which we now briefly review their definitions.

Let us consider a bounded Borel set $W \subset \mathbb{R}^{d}$ with positive measure $|W| > 0$, and let $N \in \mathbb{N}$. An $N$-BPP $X$ consists of $N$ random points $\{X_{1},\cdots,X_{N}\}$, which are independently and uniformly distributed over $W$ \cite{Lieshout_book}. Consequently, for any Borel set $B \subset W$, the probability that the number of points falling within $B$ is $\mathcal{N}(B) = k$ follows the binomial distribution: \begin{equation} P(\mathcal{N}(B) = k) = \binom{N}{k} \left(\frac{|B|}{|W|}\right)^k \left(1-\frac{|B|}{|W|}\right)^{N-k}, \quad 0 \leq k \leq N. \end{equation}

The PPP emerges as the limit of the BPP when the volume of the domain $W$ is sent to infinity while keeping the point density $\lambda = \frac{N}{|W|}$ fixed. In this limit, for a bounded Borel set $B \subset W$, the probability that $k$ points fall within $B$ converges to: \begin{equation} P(\mathcal{N}(B) = k) \rightarrow e^{-\lambda |B|} \frac{(\lambda |B|)^k}{k!}. \end{equation}

In general, a point process $X$ is a homogeneous PPP on $W$ with intensity $\lambda$ if the following two conditions are satisfied \cite{Lieshout_book}: \begin{itemize} \item The number of points $\mathcal{N}(B)$ in a bounded Borel set $B \subset W$ is a Poisson-distributed random variable with mean $\lambda |B|$. \item Given $s$ disjoint bounded Borel sets within $W$, the number of points in each set forms a collection of $s$ independent random variables. \end{itemize}

\begin{figure}[t]
%[!htb]
    \centering
    \includegraphics[width=1\linewidth]{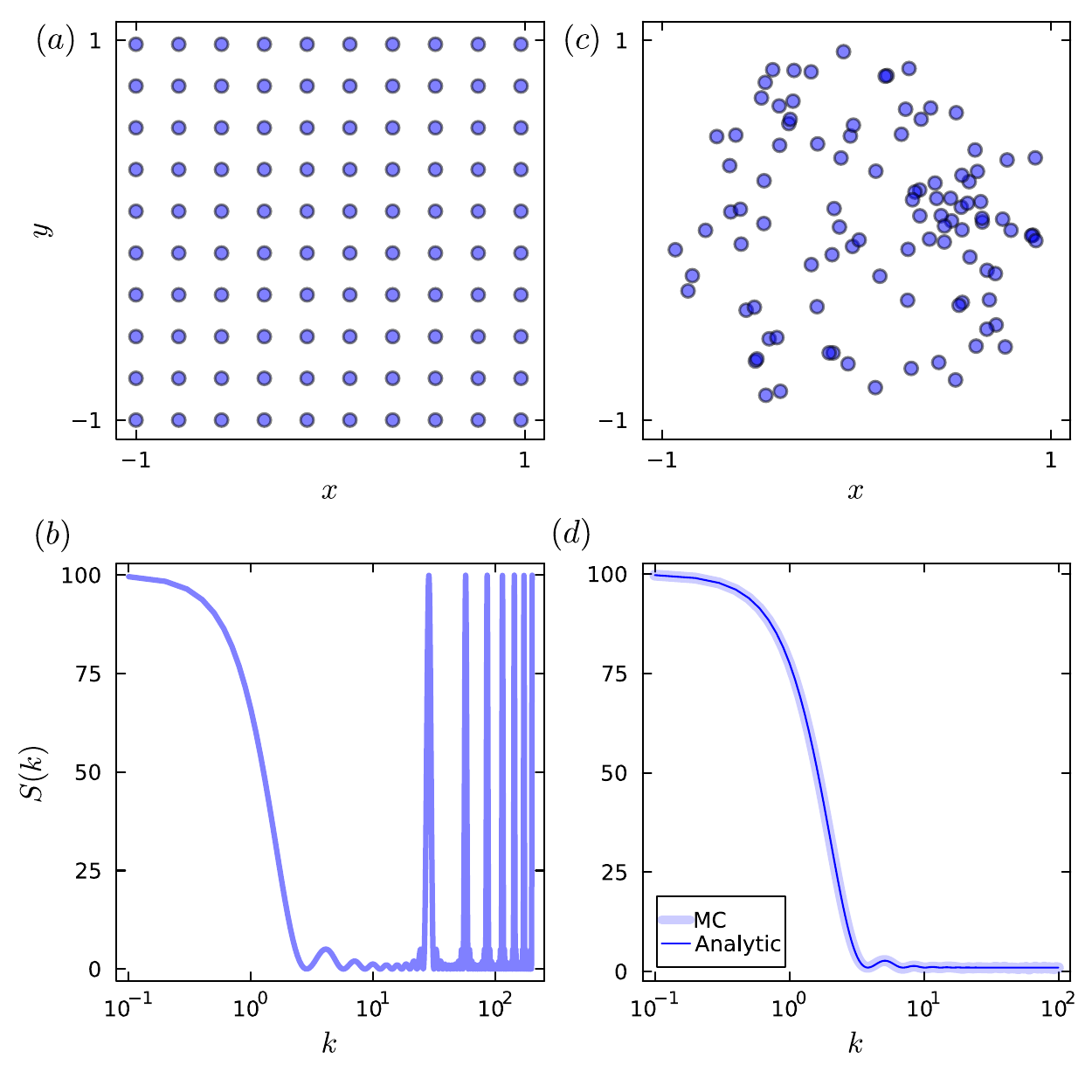}
    \caption{Structure factor of a point system on a regular lattice and BPP. Panels (a) and (b) show a system of \( N = 100 \) points on a square lattice and its corresponding structure factor \( S(\mathbf{k}) \), determined numerically and plotted along \( \mathbf{k} = (k, 0) \).
    Panels (c) and (d) depict a specific realization of a BPP with \( N = 100 \) points within a unit radius disk, and the structure factor of the corresponding process, plotted along \( \mathbf{k} = (k, 0) \). In panel (d), the thin solid line represents the analytical expression given by Eq. (\ref{S(K)_BPP}), while the thicker transparent line is obtained by numerically computing \( S(\mathbf{k}) \) using a Monte Carlo simulation, averaging over 100 realizations of the process.
}
    \label{fig:Sk_lattice_vs_PPP}
\end{figure}

Having reviewed the relevant definitions, we now consider a BPP consisting of $N$ points distributed within a disk of radius $R$ as an introductory example to demonstrate the calculation of the structure factor. In this scenario, $S(\mathbf{k})$ can be computed from the definition (\ref{def_S(k)}) as 
\begin{align}
S_{\scaleto{{\rm BPP}(N)\mathstrut}{5pt}}(\mathbf{k}) &= \frac{1}{N} \left( N + N(N-1) \left\langle e^{i\mathbf{k}\cdot(\mathbf{r}_{1}-\mathbf{r}_{2})} \right\rangle \right) \nonumber\\
&= 1 + (N-1) \iint_{\text{disk}(R)} \frac{1}{(\pi R^2)^2} e^{i\mathbf{k}\cdot(\mathbf{r}_{1}-\mathbf{r}_{2})} \, d\mathbf{r}_{1} \, d\mathbf{r}_{2} \nonumber\\
&= 1 + (N-1) \left( \frac{2J_{1}(kR)}{kR} \right)^2 \label{S(K)_BPP},
\end{align}
where $J_{1}$ is the Bessel function of the first kind, and $k$ is the modulus of the vector $\mathbf{k}$.

The structure factor of a homogeneous PPP, $S_{\scaleto{\rm{PPP}\mathstrut}{5pt}}(\mathbf{k})$, defined on the same disk, can be derived similarly. In fact, it simply involves averaging the $S_{\scaleto{\rm{BPP(N)}\mathstrut}{5pt}}(\mathbf{k})$ expression over $N$ according to the Poisson distribution.
As a result, \(S_{\scaleto{\rm{PPP}\mathstrut}{5pt}}(\mathbf{k}) \) is obtained by replacing \( N \) in Eq. (\ref{S(K)_BPP}) with the intensity $\lambda$ of the PPP.

From the previous examples, it is clear that the structure factor is a useful tool for characterizing spatial point patterns, as shown in Fig. \ref{fig:Sk_lattice_vs_PPP}. Specifically, the characteristic profile of \( S_{\scaleto{\rm{BPP(N)}\mathstrut}{5pt}}(\mathbf{k}) \), derived in (\ref{S(K)_BPP}), can serve as a benchmark for assessing the independence and uniformity of point distributions. By comparing the structure factor \( S(\mathbf{k}) \) of a generic \( N \)-point process with \( S_{\scaleto{\rm{BPP(N)}\mathstrut}{5pt}}(\mathbf{k}) \), any observed deviations can indicate whether the points of the examined process have been sampled uniformly and independently.

In a physical system, excluding the influence of an external potential, such deviations may suggest the presence of inter-particle interactions. However, for a two-dimensional point pattern, \( S(\mathbf{k}) \) is generally a two-dimensional function, making the comparison between a given structure factor and the reference somewhat inconvenient. This task becomes even more challenging as the number of spatial dimensions increases. Dimensional reduction is possible for a homogeneous point process on a spherically symmetric domain, where \( S(\mathbf{k}) \) depends only on the magnitude of \( \mathbf{k} \), thus encoding all relevant information in a single dimension. Nevertheless, the structure factor does not, in general, exhibit this symmetry.

This motivated us to introduce a modified version of the structure factor, which we call the spatial form factor (SFF), defined as
\begin{equation}\label{GFF_def_motivation}
 {\rm SFF}(k):=\operatorname{Re} \left\langle \frac{1}{N^2} \sum_{i,j=1}^{N}e^{i k|\mathbf{r}_{i}-\mathbf{r}_{j}|} \right\rangle = \left\langle \frac{1}{N^2} \sum_{i,j=1}^{N} \cos( |\mathbf{r}_{i}-\mathbf{r}_{j}|k) \right\rangle.
\end{equation}
The main difference from $S(\mathbf{k})$ is the replacement of the scalar product $\mathbf{k} \cdot (\mathbf{r}_{i} - \mathbf{r}_{j})$ with $k|\mathbf{r}_{i} - \mathbf{r}_{j}|$, where \( k \) is a scalar. Consequently, the resulting ${\rm SFF}(k)$ is a one-dimensional real function.
%complex function, of which we consider the real part.

The reader may notice the analogy between the SFF, as defined in Eq. (\ref{GFF_def_motivation}), and the spectral form factor $K(t)$. Specifically, given a matrix of dimension \( N \) with an spectrum $\sigma=\{E_i\}$ given by the set of eigenvalues $E_{i}$, $K(t)$ is defined as \cite{mehta1991random}:
\begin{equation}
    K(t) = \frac{1}{N^2} \left\langle \sum_{n,m=1}^N e^{it(E_{m}-E_{n})} \right\rangle.
\end{equation}
\begin{figure*}[t]
%[!htb]
    \centering
    \includegraphics[width=0.9\linewidth]{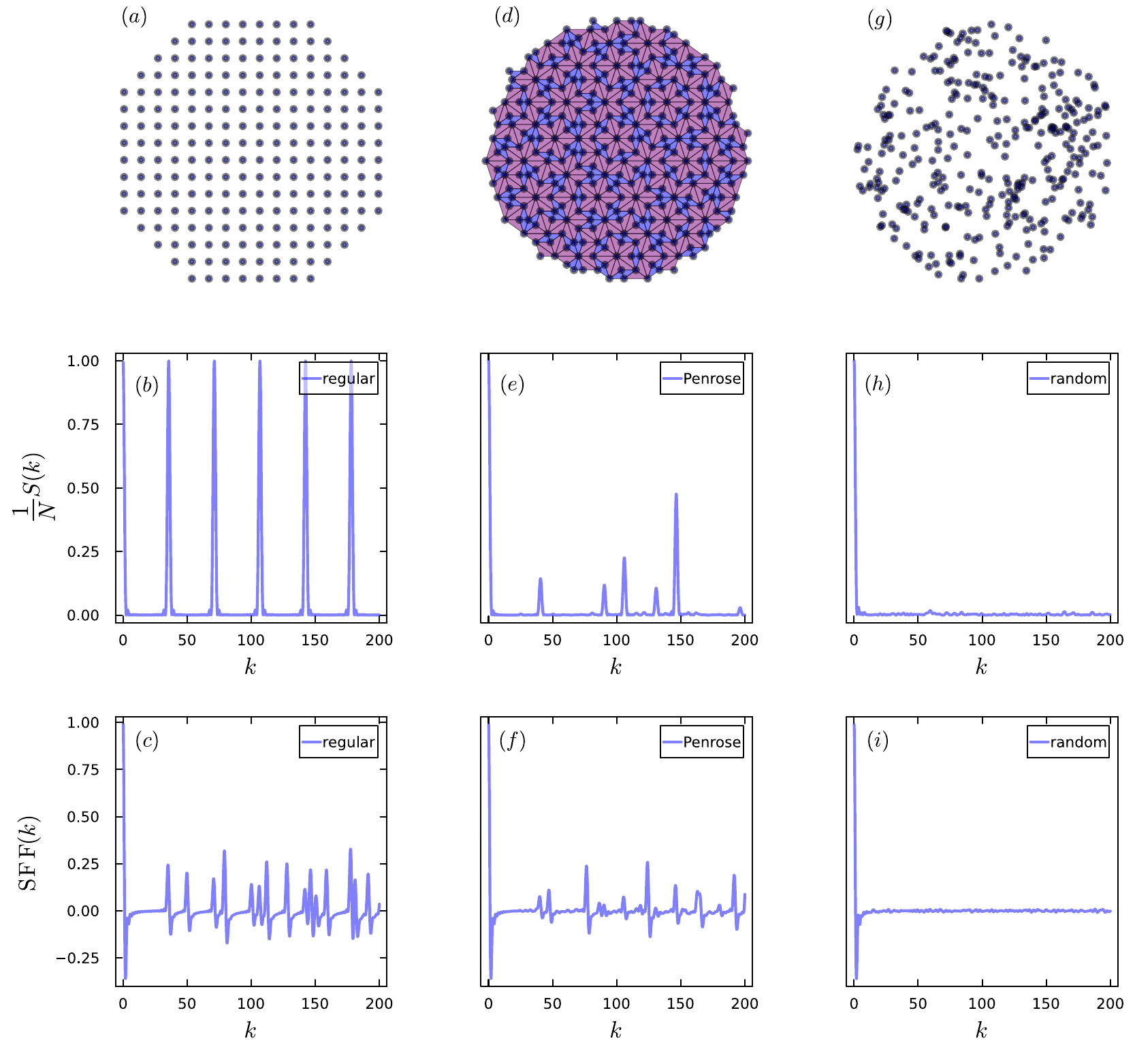}
    \caption{
Comparison of the structure factor \( S(k) \) along \(\mathbf{k} = k \hat{x}\) and the spatial form factor (SFF) for various point patterns. Panels (a), (b), and (c) illustrate a regular square lattice, its corresponding \( S(k) \), and SFF, respectively. Panels (d), (e), and (f) show a Penrose tiling (P3) with its \( S(k) \) and SFF. Panels (g), (h), and (i) present a single realization of a Poisson point process (PPP), along with its \( S(k) \) and SFF. All patterns are generated with identical numbers of points \( N \) and the same density. The structure factor signals crystalline or regular order through the presence of spikes. By contrast, the SFF provides complementary information regarding the spacing distribution and is suited for the characterization of disordered systems.
}
    \label{fig:Penrose_random_regular_DFF}
\end{figure*}
The spectral form factor has been used to study the properties of energy levels in various physical systems, including atomic nuclei, quantum chaotic systems, and black holes \cite{Leviandier86,WilkieBrumer91,Alhassid93,Ma95,Haake2010,Cotler17,delcampo17,shir2023range}. In this context, the spacing between such eigenvalues plays a crucial role, e.g., in assessing integrability and its breakdown. 
We note that  $K(t)$ is a real function that can be written as  
\begin{equation}
\label{sFFcos}
    K(t) = \frac{1}{N^2}\left\langle  \sum_{n,m=1}^N \cos(|E_{m}-E_{n}|t) \right\rangle,
\end{equation}
where we have introduced the absolute value to emphasize that $K(t)$ depends on all eigenvalue spacings $|E_{i} - E_{j}|$. Equation (\ref{sFFcos}) reflects that the SFF (\ref{GFF_def_motivation}) is a natural generalization of $ K(t)$  where a spatial point pattern replaces the role of the Hamiltonian spectrum. 
In the following section, after presenting the general formal definition of the SFF for a generic spatial point process, we provide its explicit calculation for the BPP and PPP.

\section{Definition and General Properties of the Distance Form Factor}\label{SecGFFdef}
\subsection{Definition and numerical examples}

In a generic spatial point process defined over a domain $\mathcal{D}$, an event $E$ corresponds to a specific arrangement of $N$ points within $\mathcal{D}$:
\begin{equation}
    E=\{\mathbf{r}_{1},\cdots,\mathbf{r}_{N} \} \quad \text{,} \quad \mathbf{r}_{i} \in \mathcal{D} \quad \forall i.
\end{equation}
We define the spatial form factor of the process as 
\begin{equation}\label{sff_def}
   {\rm SFF}(k)=\frac{1}{N^2} \biggl< \sum_{i,j=1}^{N} \cos\left(|\mathbf{r}_{i}-\mathbf{r}_{j}|k\right) \biggr>,
\end{equation}
where the average $\langle\cdot\rangle$ is performed over all the possible configurations $E$ of the point system. In particular, denoting by $\rho(E):=\rho(\mathbf{r}_{1},\cdots, \mathbf{r}_{N})$ the probability density of a certain configuration $E$, 
\begin{equation}
    {\rm SFF}(k)=\frac{1}{N^2}\sum_{i,j=1}^{N}\left(\int_{\mathcal{D}^{N}} \cos(|\mathbf{r}_{i}-\mathbf{r}_{j}|k) \rho(\mathbf{r}_{1},\cdots, \mathbf{r}_{N}) d^{N}\mathbf{r}\right),
\end{equation}
where $d^{N}\mathbf{r}=d\mathbf{r}_{1} \cdots d\mathbf{r}_{N}$.
The SFF can be used to characterize different classes of point patterns, with its behavior varying according to the salient feature in each class. By way of example, in Fig. \ref{fig:Penrose_random_regular_DFF}, the case of a regular square lattice pattern is compared with that of a  Penrose tiling as an example of a quasiperiodic pattern \cite{Penrose79,Grunbaum87} and a homogeneous PPP. In all cases, the SFF exhibits a decay from unit value until a dip and a subsequent steep increase towards a plateau. The presence of spikes in the plateau is characteristic of the regular pattern, which is reduced in the quasiperiodic case and washed out in the random case. It is this latter case that motivates the introduction of the SFF, given its applications to quantum chaos and defect characterization already discussed.

\subsection{Relation to the spacing distribution}

By separating the terms in Eq. (\ref{sff_def}) into the cases with $i=j$ and $i\neq j$ and applying the linearity of the expectation value, the SFF can be written as
\begin{eqnarray}
{\rm SFF}(k) &=& \frac{1}{N^2}\left(N + \sum_{(i \neq j)=1}^{N} \left< \cos(|\mathbf{r}_{i} - \mathbf{r}_{j}|k) \right> \right) \\
&=& \frac{1}{N} + \left(1 - \frac{1}{N}\right) \left< \cos(|\mathbf{r}_{1} - \mathbf{r}_{2}|k) \right> \label{eq:sff_before_computing_average},
\end{eqnarray}
where the last line follows from assuming permutation symmetry in the indices.
It is then convenient to consider the two-point distance, $s:=|\mathbf{r}_{1}-\mathbf{r}_{2}|$, in terms of which   
\begin{equation}
    \biggl<\cos(|\mathbf{r}_{1}-\mathbf{r}_{2}|k) \biggr>=\biggl< \cos(sk) \biggr>=\int_{0}^{s_{max}}\cos(sk)P_{\mathcal{D}}(s)\,ds ,
    \label{sffcomputationdD}
\end{equation}
where $P_\mathcal{D}(s)$ represents the probability distribution for the distance $s$ between two points on the domain $\mathcal{D}$, with $s_{max}$ its maximum value. This transformation reduces the dimension of our problem from $d$ to 1 by introducing the probability density $P_\mathcal{D}(s)$.
%, which is no longer uniform.

Thus, the SFF is related to the even Fourier transform of $P_\mathcal{D}(s)$ as 
\begin{align}
\label{eqDFFPs}
{\rm SFF}(k) &=\frac{1}{N} + \left(1 - \frac{1}{N}\right) \int_{0}^{s_{max}}\cos(sk)P_{\mathcal{D}}(s) \,ds \nonumber\\
&=\frac{1}{N}+\left(1-\frac{1}{N}\right)\mathcal{I}_\mathcal{D}(k),
\end{align}
where we have defined 
\begin{equation}\label{def_I_d}
\mathcal{I}_\mathcal{D}(k) := \int_{0}^{s_{max}}\cos(sk)P_{\mathcal{D}}(s)\,ds, \end{equation}

Conversely, given the SFF, the expression for the probability $P_\mathcal{D}(s)$ can be retrieved as follows. Let us
multiply both sides of Eq. (\ref{def_I_d}) by $\cos(s^{\prime}k)$ and integrate over $k$ to find
\begin{equation}\label{invert_I_d}
    \int_{-\infty}^{\infty}\mathcal{I}_\mathcal{D}(k)\cos(s^{\prime}k) \, dk =\int_{-\infty}^{\infty}\int_{0}^{s_{max}}\cos(s^{\prime}k)\cos(sk)P_\mathcal{D}(s) \, dk ds.
\end{equation}
Using the result
\begin{equation}
    \int_{-\infty}^{\infty}\cos(s^{\prime}k)\cos(sk) dk=\pi(\delta(s+s^{\prime})+\delta(s-s^{\prime})),
\end{equation}
and, assuming that $0\leq s^{\prime}\leq s_{max}$, Eq. (\ref{invert_I_d}) becomes
\begin{equation}
    \int_{-\infty}^{\infty}\mathcal{I}_\mathcal{D}(k)\cos(s^{\prime}k) \, dk = \pi P_\mathcal{D}(s^{\prime}).
\end{equation}
Therefore, given the $\mathrm{SFF(k)}$, the inverse transformation to find $P_\mathcal{D}$ is
\begin{equation}
\label{eqPsDFF}
    P_\mathcal{D}(s)=\frac{N}{\pi(N-1)}\int_{-\infty}^{\infty}\left(\mathrm{SFF}(k)-\frac{1}{N}\right)\cos(sk) \,dk.
\end{equation}
Equations (\ref{eqDFFPs}) and (\ref{eqPsDFF}) thus establish a direct relation between the SFF and the spacing distribution. 
%Although our primary focus is on applying the SFF to stochastic point processes, 
%As a result, 

\section{Spatial Form Factor for Binomial and Poisson Point Processes on a $d$-Dimensional Ball}\label{GFFPPP}

In the following, we focus on the N-BPP defined on a $d$-dimensional ball of radius $R$, $B_d(R)=\{ \mathbf{r} \in \mathbb{R}^{d}: ||\mathbf{r}||\leq R   \}\ $.
Extending our results to the homogeneous PPP or cases where $N$ fluctuates according to a generic distribution is straightforward, as it simply requires averaging the results obtained for the BPP case over $N$.

Under these assumptions, due to the independence between random points, all the configurations $E$ are equally likely, with a corresponding probability density fixed by the normalization condition:
\begin{equation} \label{eq:Prob_density_PPP}
    \rho(E)=\left(\frac{1}{V_{B_d(R)}}\right)^{N},
\end{equation}
where $V_{B_d(R)}$ is the volume of the $d$-dimensional ball:
\begin{equation}
V_{B_d(R)}=\frac{\pi^{\frac{d}{2}}}{\Gamma(\frac{d}{2}+1)} R^{d}.
\end{equation}
We now derive the analytical expression of the SFF in this scenario by explicitly evaluating the average in Eq. (\ref{sff_def}).

The remaining average in Eq. (\ref{eq:sff_before_computing_average}) can be computed by integrating over all possible system configurations, uniformly weighted according to Eq. (\ref{eq:Prob_density_PPP}). This is expressed as
\begin{align}
\left<\cos(|\mathbf{r}_{1}-\mathbf{r}_{2}|k)\right> &= \left(\frac{1}{V_{B_d(R)}}\right)^N \int_{B_d(R)} \cos(|\mathbf{r}_{1}-\mathbf{r}_{2}|k) \, d^{N}\mathbf{r} \nonumber\\
&= \left(\frac{1}{V_{B_d(R)}}\right)^2 \int_{B_d(R)} \cos(|\mathbf{r}_{1}-\mathbf{r}_{2}|k) \, d\mathbf{r}_{1} \, d\mathbf{r}_{2}.
\end{align}
However, it reveals more convenient to make use of Eq. (\ref{sffcomputationdD}), identifying the corresponding $P_{\mathcal{D}}(s)$ with  the probability distribution $P_{d}(s)$ for the distance $s$ between two independent and uniformly distributed random points on a $d$-dimensional ball.
The expression for $P_{d}(s)$ is known from the  Ball Line Picking problem in stochastic geometry, and it is given by \cite{kendallgeoprob,Santalointgeo}
\begin{equation} \label{Psball_linepicking}
    P_{d}(s)=\frac{d \cdot s^{d-1}}{R^d}I_{1-\frac{s^2}{4R^2}}\left(\frac{d+1}{2},\frac{1}{2}\right),
\end{equation}
where $I_{z}(a,b)$ is the regularized Beta function, defined in terms of the incomplete beta function $B(z;a,b)$ and the beta function $B(a,b)$ as \cite{Gradshteyn96}
\begin{equation} \label{eq:def_beta_regularized}
    I_{z}(a,b):=\frac{B(z;a,b)}{B(a,b)}.
\end{equation}
We recall the integral representation of $B(z;a,b)$, given by
\begin{equation} \label{repbetaincomplete}
    B(z;a,b)=\int_{0}^{z} u^{a-1} (1-u)^{b-1} \,du.
\end{equation}
Furthermore, $B(a,b)$ can be expressed in terms of the Gamma function as 
\begin{equation} \label{betarepgamma}
    B(a,b)=\frac{\Gamma(a)\Gamma(b)}{\Gamma(a+b)}.
\end{equation}
By making use of these identities, 
%equations %(\ref{eq:def_beta_regularized}), (\ref{repbetaincomplete}) and (\ref{betarepgamma}), 
we can write $P_{d}(s)$ in the form

\begin{equation} \label{Ps_integral_representation}
    P_{d}(s)=\frac{d \cdot s^{d-1}}{R^{d}}\frac{\Gamma(\frac{d}{2}+1)}{\Gamma(\frac{d+1}{2})\Gamma(\frac{1}{2})}\int_{0}^{1-\frac{s^2}{4R^2}}u^{\frac{d-1}{2}}(1-u)^{-\frac{1}{2}}du.
\end{equation}
After plugging Eq. (\ref{Ps_integral_representation}) into Eq. (\ref{def_I_d}), one is left to evaluate the integral
\begin{align} 
\mathcal{I}_{d}(k) &:=  \int_{0}^{2R}\cos(sk)P_{d}(s) \, ds \nonumber \\
&=\frac{d}{R^{d}}\frac{\Gamma(\frac{d}{2}+1)}{\Gamma(\frac{d+1}{2})\Gamma(\frac{1}{2})}\nonumber\\
&\times\int_{0}^{2R}\cos(sk) s^{d-1} \left[\int_{0}^{1-\frac{s^2}{4R^2}}u^{\frac{d-1}{2}}(1-u)^{-\frac{1}{2}} \, du\right] ds.\label{Id_2}
\end{align}
To ease the notation, let $g(s)$ represent the integral inside the square bracket in Eq. (\ref{Id_2}),
\begin{equation}
    g(s):= \int_{0}^{1-\frac{s^2}{4R^2}}u^{\frac{d-1}{2}}(1-u)^{-\frac{1}{2}} \, du,
\end{equation}
so that we can more compactly write 
\begin{equation}
     \mathcal{I}_{d}(k)=\frac{d}{R^{d}}\frac{\Gamma(\frac{d}{2}+1)}{\Gamma(\frac{d+1}{2})\Gamma(\frac{1}{2})} \int_{0}^{2R} \cos(sk)s^{d-1}g(s)\,ds.
\end{equation}
We will now proceed by integration by parts. Let us first denote with $G(s)$ the primitive of $s^{d-1}\cos(sk)$, which can be computed explicitly in terms of the hypergeometric function ${}_1F_2$ \cite{Gradshteyn96,NIST:DLMF} as
\begin{equation}\label{Gofs}
    G(s):= \int_{0}^{s}x^{d-1}\cos(xk)\,dx=\frac{s^d}{d} \, {}_1F_2\left(\frac{d}{2}; \frac{1}{2}, 1 + \frac{d}{2}; -\frac{1}{4} k^2 s^2\right).
\end{equation}
Integrating $\mathcal{I}_{d}$ by parts then yields
\begin{align}
\mathcal{I}_{d} &= \frac{d}{R^{d}}\frac{\Gamma(\frac{d}{2}+1)}{\Gamma(\frac{d+1}{2})\Gamma(\frac{1}{2})} \left( \left[ G(s)g(s)\right]\big|_{0}^{2R}-\int_{0}^{2R}G(s)g^{\prime}(s) \, ds \right) \nonumber \\
&= -\frac{d}{R^{d}}\frac{\Gamma(\frac{d}{2}+1)}{\Gamma(\frac{d+1}{2})\Gamma(\frac{1}{2})}\int_{0}^{2R}G(s)g^{\prime}(s) \, ds, \label{I_d_after_int_by_parts}
\end{align}
where, to get to the last line, we used $G(0)=0$ and $g(2R)=0$.
Furthermore, the derivative of $g(s)$ can be evaluated via the fundamental theorem of calculus, 
\begin{equation}\label{theorem_calculus}
    \frac{d}{dx}\left(\int_{0}^{h(x)} f(u) du\right)=f(h(x)) h^{\prime}(x).
\end{equation}
In particular,
\begin{equation}\label{derivativeg}
    g^{\prime}(s)
    =-\frac{1}{R}\left(1-\frac{s^{2}}{4R^2}\right)^{\frac{d-1}{2}}.
\end{equation}
We can then plug the explicit expressions for $G(s)$ and $g^{\prime}(s)$, respectively given in Eq. (\ref{Gofs}) and Eq. (\ref{derivativeg}), into Eq. (\ref{I_d_after_int_by_parts}) and perform the remaining integration to find
\begin{align}
&\mathcal{I}_{d} = \frac{1}{R^{d+1}} \frac{\Gamma\left(\frac{d}{2}+1\right)}{\Gamma\left(\frac{d+1}{2}\right)\Gamma\left(\frac{1}{2}\right)}\nonumber\\
& \qquad\times\int_{0}^{2R} s^d \, {}_1F_2\left(\frac{d}{2}; \frac{1}{2}, 1 + \frac{d}{2}; -\frac{1}{4} k^2 s^2\right) \left(1 - \frac{s^{2}}{4R^2}\right)^{\frac{d-1}{2}} ds \nonumber\\
&= \sqrt{\pi} (d!) \Gamma\left(\frac{d}{2}+1\right) {}_2\tilde{F}_3\left(\frac{d+1}{2}, \frac{d}{2}; \frac{1}{2}, \frac{d+2}{2}, d+1; -k^{2}R^{2}\right), \label{Id_after_integration}
\end{align}
where we used the following properties of the Gamma function:
\begin{equation}\label{gammaproperties}
    \Gamma\left(\frac{1}{2}\right)=\sqrt{\pi} \quad \text{,} \quad \Gamma(d) \cdot d=d! .
\end{equation}

The resulting expression for $\mathcal{I}_{d}$ in (\ref{Id_after_integration}) can be further simplified by recalling the definition of the hypergeometric regularized function ${}_2\tilde{F}_3$ in terms of the standard $_2F_3$ \cite{Gradshteyn96}, 
\begin{equation}
{}_2\tilde{F}_3(a_{1},a_{2};b_{1},b_{2},b_{3};z) := \frac{{}_2F_3(a_{1},a_{2};b_{1},b_{2},b_{3};z)}{\Gamma(b_{1})\Gamma(b_{2})\Gamma(b_{3})}.
\end{equation}
Therefore,
\begin{align}
\mathcal{I}_{d} &= \frac{\sqrt{\pi}(d !) \Gamma\left(\frac{d}{2}+1\right)}{\Gamma\left(\frac{1}{2}\right)\Gamma\left(\frac{d+2}{2}\right)\Gamma\left(d+1\right)}{}_2F_3\left(\frac{d+1}{2},\frac{d}{2};\frac{1}{2},\frac{d+2}{2},d+1;-k^{2}R^{2}\right) \nonumber \\
&= {}_2F_3\left(\frac{d+1}{2},\frac{d}{2};\frac{1}{2},\frac{d+2}{2},d+1;-k^{2}R^{2}\right), \label{eq:final_I_d}
\end{align}
where we made again use of the properties (\ref{gammaproperties}).

Finally, by plugging the result of Eq. (\ref{eq:final_I_d}) in Eq. (\ref{eqDFFPs}), the expression for the SFF reads
\begin{equation}\label{finalsimplifiedSFFdBall}
{\rm SFF}(k)=\frac{1}{N}+\left(1-\frac{1}{N}\right) {}_2F_3\left(\frac{d+1}{2},\frac{d}{2};\frac{1}{2},\frac{d+2}{2},d+1;-k^{2}R^{2}\right),
\end{equation}
which constitutes our main result for the BPP. As previously mentioned, extending Eq. (\ref{finalsimplifiedSFFdBall}) to cases where $N$ fluctuates across different trials according to a generic distribution is straightforward and involves averaging the expression (\ref{finalsimplifiedSFFdBall}) over $N$.
Special care is only required when dealing with the case \( N = 0 \), as the SFF is not well-defined in this scenario.
To address this, we conventionally assign a value of 0 to the \({\rm SFF}\) when \( N = 0 \). In particular, for a homogeneous PPP we have:
\begin{align}
&{\rm SFF_{\scaleto{\rm{PPP}\mathstrut}{5pt}}}(k)=
f(\lambda)+\left(1-f(\lambda)\right)\nonumber\\
&\qquad \quad \times{}_2F_3\left(\frac{d+1}{2},\frac{d}{2};\frac{1}{2},\frac{d+2}{2},d+1;-k^{2}R^{2}\right)
,
\end{align}
where $\lambda$ is the intensity of the PPP over $B_d(R)$ and
\begin{equation}
    f(\lambda)=e^{-\lambda}\left(-\gamma+\text{Chi}(\lambda)-\log(\lambda)+\text{Shi}(\lambda)\right),
\end{equation}
with $\text{Chi}(\lambda)$ and $\text{Shi}(\lambda)$ being respectively the hyperbolic cosine and sine integrals \cite{Gradshteyn96}, and $\gamma$ the Euler-Mascheroni constant. 

In the next subsection, we consider general features of the SFF, as well as some special cases of Eq. (\ref{finalsimplifiedSFFdBall}) and discuss connections with random matrix theory.   

\begin{table*}[t]
\centering
\makegapedcells
\setlength\tabcolsep{6pt}
    \begin{tabularx}{\linewidth}{l >{\centering\arraybackslash}X} % Left-align first column, center second column
   \Xhline{0.8pt}
   & $\displaystyle \mathcal{I}_d(k)$ \\ % Add an empty first column to align \mathcal{I}_d with the second column
   \Xhline{0.1pt} % Thin line below the first row
$d=1$ & $\displaystyle \frac{\sin^{2}(kR)}{(kR)^2}$ \\ % Left-align d=1, center the formula
$d=2$ & $\displaystyle \frac{-2kR(1 + 2J_0(2kR)) + 6J_1(2kR)}{k^3R^3}$ \\ % Left-align d=2, center the formula
$d=3$ & $\displaystyle -\frac{9 \left(3 k^2 R^2+2 k R \left(k^2 R^2-5\right) \sin (2 k R)+\left(7 k^2 R^2-5\right) \cos (2 k R)+5\right)}{2 k^6 R^6}$ \\ % Left-align d=3, center the formula
   \Xhline{0.8pt} % Thick line at the bottom
\end{tabularx}
\caption{The integral $\mathcal{I}_d(k)$ determines the spatial form factor of a BPP according to the identity ${\rm SFF}(k)=\frac{1}{N}+\left(1-\frac{1}{N}\right)\mathcal{I}_d(k)$.  The explicit form of $\mathcal{I}_d(k)$ is quoted for $d=1,2,3$.}\label{tab:I_d_1_2_3}
\end{table*}

\begin{figure*}[t]
    \centering
    \includegraphics[width=0.8\linewidth]{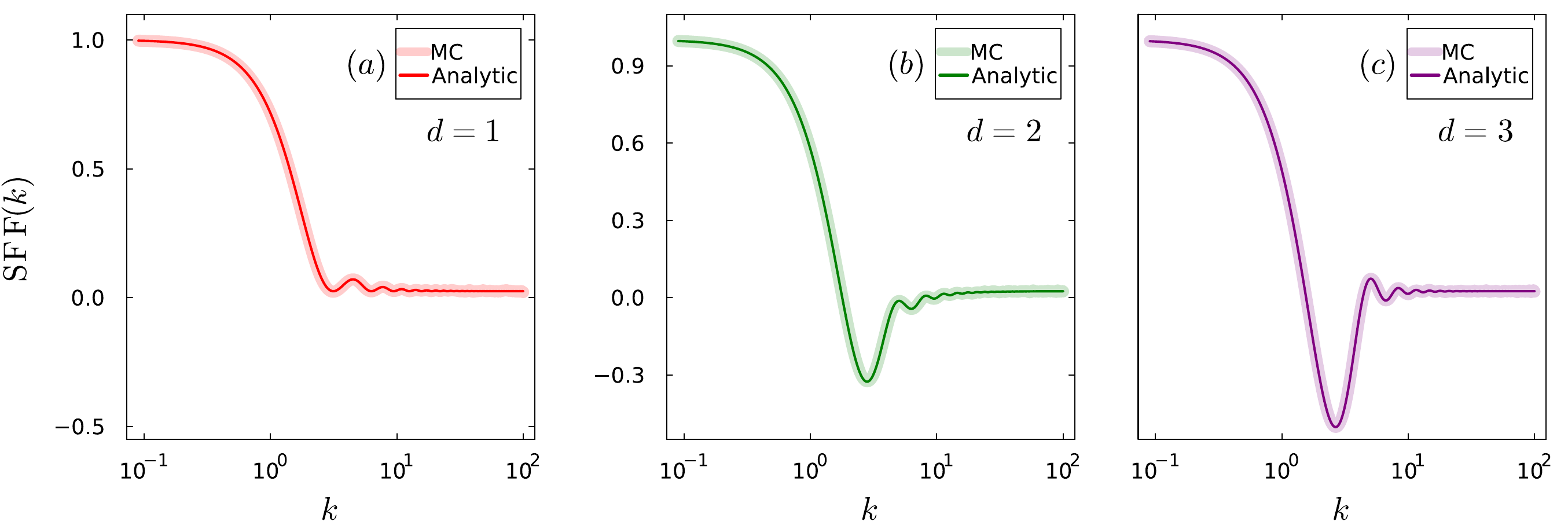}
    \caption{Spatial form factor of a BPP on a $d$-dimensional ball with unit radius. Panels (a), (b), and (c) correspond respectively to $d=1,2,3$.
    In each of the three cases, the solid thin line represents the analytic expression given in Eq. (\ref{finalsimplifiedSFFdBall}), while the transparent thicker line corresponds to the SFF computed numerically in a Monte Carlo simulation with $N=40$ points, averaging over 300 randomly generated configurations.}
    \label{fig:sff_d1_d2_d3_analytic_MC}
\end{figure*}

\section{Characterization of the Spatial Form Factor for Stochastic Point Processes}\label{SecCharGFF}
\subsection{Asymptotics, special cases and connections to quantum chaos}

A Taylor series expansion of the SFF (\ref{finalsimplifiedSFFdBall}) for a BPP around $kR=0$ yields
\begin{equation}
    {\rm SFF}(k)=1-\frac{(N-1)d}{N(d+2)}(kR)^2+\frac{d(d+3)(N-1)}{6N(d+4)(d+2)}(kR)^4+\mathcal{O}(k^6R^6),
\end{equation}
indicating a quadratic decay from unit value for small values of $kR$ that is interrupted around
\begin{equation}
    kR\approx\sqrt{6+\frac{6}{3+d}},
\end{equation}
that is independent of $N$.
Furthermore, from the Riemann–Lebesgue lemma, it follows that $\mathcal{I}_{d}$ vanishes for large values of $kR$, so that
\begin{equation}
    {\rm SFF}(k)\sim \frac{1}{N}.
\end{equation}
Thus, the SFF is expected to exhibit a decay from unit value, forming a dip, with subsequent growth leading to a plateau, as anticipated by the numerical examples in Sec. \ref{SecGFFdef}. 

The special cases of our formula (\ref{finalsimplifiedSFFdBall}) for dimensions 1, 2, and 3 are explicitly presented in Table \ref{tab:I_d_1_2_3}, with the corresponding plots shown in Fig. \ref{fig:sff_d1_d2_d3_analytic_MC}.

What stands out is the characteristic structure of sub-figures \ref{fig:sff_d1_d2_d3_analytic_MC}(b) and (c), consisting of a dip followed by a ramp and a subsequent plateau. The observed behavior is consistent with the short and long $k$ asymptotic.

The behavior of the SFF is reminiscent of that in the spectral form factor in quantum chaotic systems. Quantum chaos manifests in correlations between the eigenvalues of an isolated Hamiltonian quantum system \cite{Haake2010}. Its characterization often relies on the eigenvalue spacing distribution. The absence of degeneracies in the energy spectrum leads to a suppression of the probability of finding a pair of eigenvalues at vanishing spacing, known as the correlation hole. Spectral signatures can be conveniently revealed by analyzing the spectral form factor, which is the average of the absolute value of the Fourier transform of the eigenvalue density \cite{Leviandier86,WilkieBrumer91,Alhassid93,Ma95}.
The spectral form factor exhibits a short-time decay form unit value, forming a dip and a subsequent ramp before saturating at a plateau  \cite{Haake2010,Cotler17,delcampo17}. 

It might appear surprising that the SFF of a BPP, where the random points are uncorrelated, shares this behavior. However, this is purely a geometrical effect due to the dimension of the domain space, which introduces a fictitious repulsion between points. In fact, as the space dimension increases, it becomes increasingly rare to sample points close to each other. 
This heuristic statement is mathematically encoded in the nearest neighbor spacing distribution. For a BPP on a $d$-dimensional ball, this distribution is given by \cite{Haake2010}
\begin{equation}
     P^{\text{\tiny NN}}_{d}(S)=\alpha d S^{d-1} e^{-\alpha S^{d}},\label{eq:NN_spacing_PPP_d_ball}
\end{equation}
with further details discussed in Sec. \ref{sec:GFF_decomposition}.
For $d=2$, it simplifies to the well-known Wigner-Dyson distribution:
\begin{equation}
    P^{\text{\tiny NN}}_{2}(S)=\frac{\pi}{2}Se^{-\frac{\pi^2}{4}S^2}.
\end{equation}
This distribution is ubiquitous in the study of quantum chaos, where $S$ denotes the eigenvalue spacing \cite{mehta1991random,Haake2010,Forrester2010}, and also characterizes the spatial patterns of topological defects generated in crossing a continuous phase transition \cite{delcampo22,thudiyangal2024}.

In the limit $S \rightarrow 0$, for $d>1$,  (\ref{eq:NN_spacing_PPP_d_ball}) goes to zero, indicating the presence of an effect analogous to the energy level repulsion in the spectral statistics of disordered quantum systems, which leads to the presence of the correlation hole.
This explains why this feature is absent in Fig. \ref{fig:sff_d1_d2_d3_analytic_MC}(a), corresponding to the 1$d$ case. In fact, in that scenario, the nearest neighbor spacing distribution is Poissonian, 
\begin{equation}
 P^{\text{\tiny NN}}_{1}(S)=e^{-S}, 
\end{equation}
which does not vanish in the small spacing limit, resulting in the absence of the correlation hole.

\subsection{Spatial form factor decomposition}\label{sec:GFF_decomposition}

The SFF contains information on the spacing between any pair of points of the process. 
It is in fact, up to constant factors, the Fourier transform of the spacing distribution $P_{d}(s)$, 
as it is apparent from Eq. (\ref{eqDFFPs}).
However, it is possible to separate the contributions to ${\rm SFF}(k)$ based on the neighborhood degree between pairs of points. This is analogous to the decomposition recently put forward in the canonical spectral form factor in the context of quantum chaos \cite{shir2023range}. In particular, we have
\begin{equation}
    {\rm SFF}(k)=\frac{1}{N}+\sum_{n=1}^{N-1}{\rm SFF}^{(n)}(k),
\end{equation}
where  ${\rm SFF}^{(n)}(k)$ denotes the spatial form factor computed considering only $n$-th nearest neighboring points. Specifically, 
\begin{equation}\label{N-th_Gff}
{\rm SFF}^{(n)}(k):=\frac{1}{N^2} \biggl< \sum_{i=1}^{N} \cos\left(|\mathbf{r}_{i}-\mathbf{r}_{n^{\text{th}}(i)}|k\right) \biggr>,
\end{equation}
with $\mathbf{r}_{n^{\text{th}}(i)}$ representing the position of the $n$-th closest point to the point at $\mathbf{r}_i$.

While there is an analogy with the decomposition of the spectral form factor, we emphasize an important difference here. In the spectral form factor, the $n$-th nearest neighbor of the $i$-th energy level is defined as the $(i+n)$-th level, assuming the levels are ordered, with their spacing given by $s_{i}^{(n)} = E_{i+n} - E_{i}$ \cite{shir2023range}. 
In contrast, for the spatial form factor, the energy levels are replaced by the positions of the points. Thus, although the neighbor definition used in the spectral form factor can still be applied in a one-dimensional point process, the natural ordering that makes this possible is lost in higher dimensions.

\begin{figure*}[t]
    \centering
\includegraphics[width=1\linewidth]{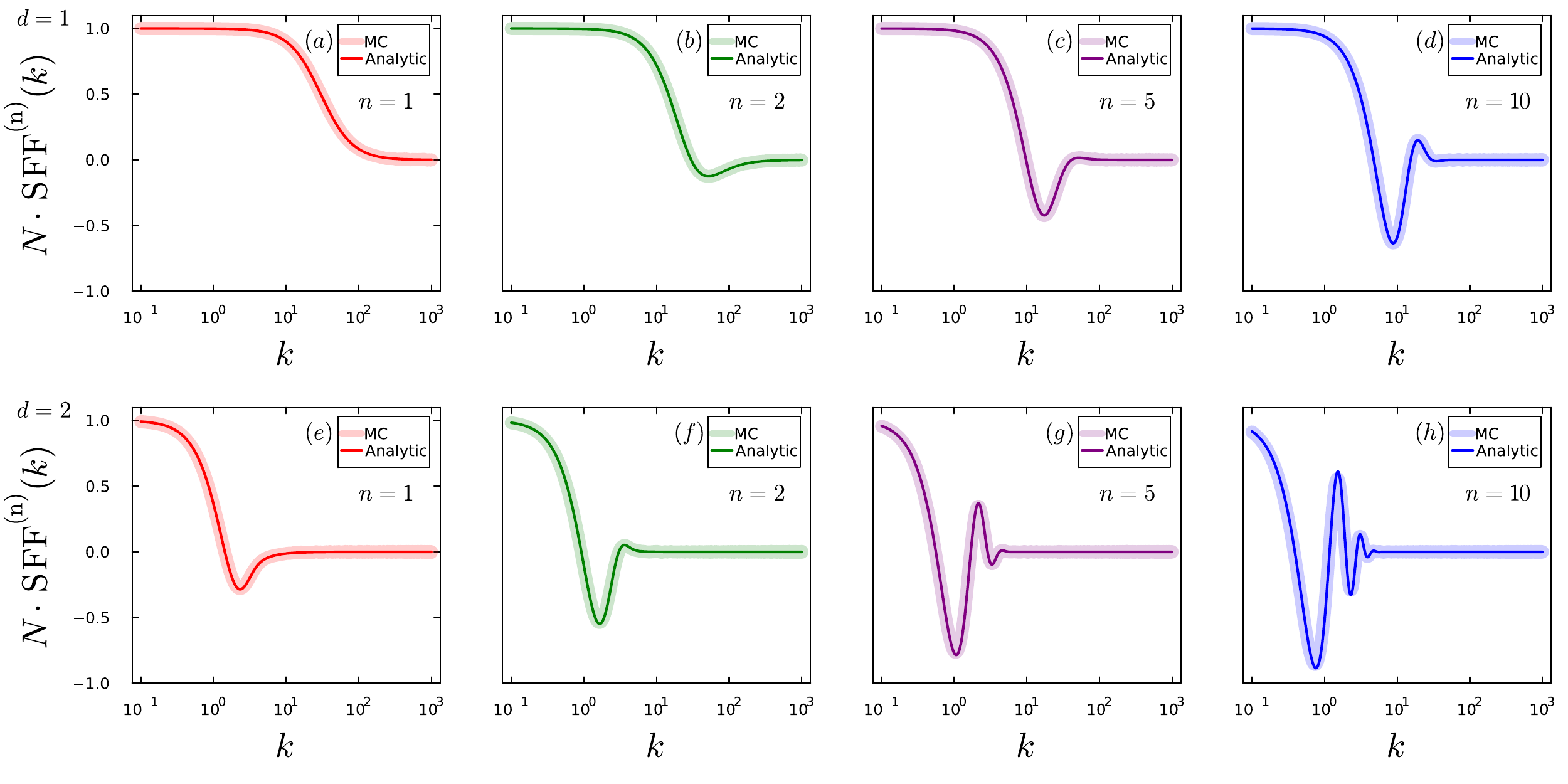}
    \caption{Different components of the \(\mathrm{SFF}\) for a BPP of \(N=1500\) points.
Panels (a), (b), (c) and (d) correspond respectively to the first, second, fifth and tenth component of the \(\mathrm{SFF}\) for the 1D process on a segment of length \(2R=100\).
Panels (e), (f), (g) and (h) show the same components of the \(\mathrm{SFF}\) for the 2D process on a disk of radius \(R=50\).
In all cases, the thin solid line corresponds to the analytical expression provided in Table (\ref{tab:gff_components}), while the thick transparent line corresponds to the \(\mathrm{SFF}^{(n)}\) computed numerically by a Monte Carlo simulation, averaging over $300$ trials.
For convenience, in all the panels, we have plotted the rescaled \(\mathrm{SFF}^{(n)}\) with the total number of points.
}
    \label{fig:gff_decomposition}
\end{figure*}

Equation (\ref{N-th_Gff}) can be further rewritten as
\begin{equation}\label{GFF_decompose_integral_form}
 {\rm SFF}^{(n)}(k)=\frac{1}{N}\int_{0}^{2R}\cos(sk)P_{d}^{(n)}(s)\,ds, 
\end{equation}
with $P_{d}^{(n)}$ the $n$-th nearest neighbor spacing probability density. In the limit where the number of points $N\rightarrow \infty$, and provided that we rescale the spacing $s$ with the mean $n$-th nearest neighbor spacing distance $\langle s^{(n)}\rangle$ ($s\rightarrow \frac{s}{\langle s^{(n)}\rangle}=:S$), such probability assumes the simple form \cite{Haake2010}
\begin{equation}\label{P_{d}^{(l)}(s)}
    P_{d}^{(n)}(s)\,ds\rightarrow P_{d}^{(n)}(S)\,dS=\frac{\alpha^{n}}{(n-1)!}S^{(dn-1)}d e^{-S^{d}\alpha} \,dS,
\end{equation}
where $\alpha:=\frac{\Gamma^{d}(\frac{1}{d}+n)}{\Gamma^{d}(n)}$, and the explicit expression for $\langle s^{(n)}\rangle$ is given by
\begin{equation}\label{mean_s^{(l)}}
\langle s^{(n)}\rangle=\frac{R\Gamma(\frac{1}{d}+n)}{N^{\frac{1}{d}}\Gamma(n)}.
\end{equation}
Using Eq. (\ref{P_{d}^{(l)}(s)}) we can then compute the large $N$ limit of ${\rm SFF}^{(n)}(k)$, which reads 
\begin{align}
&{\rm SFF}^{(n)}(k) = \frac{1}{N}\int_{0}^{\infty} \cos\left(\langle s^{(n)}\rangle Sk\right) P_{d}^{(n)}(S) \, dS \nonumber\\
&= \frac{\alpha^{n} d}{N(n-1)!}\int_{0}^{\infty} \cos\left(\frac{\rho \Gamma\left(\frac{1}{d} + n\right)}{\Gamma(n)} Sk\right) S^{(dn-1)} e^{-S^{d}\alpha} \, dS \nonumber\\
&= \frac{d}{N(n-1)!}\int_{0}^{\infty} \cos(\rho k x) x^{(dn-1)} e^{-x^{d}} \, dx, \label{calculation_GFF_l-th-thermodynamic_limit}
\end{align}
where we have defined the ratio $\rho:=\frac{R}{N^{\frac{1}{d}}}$, and, in the last line, we have performed the change of variables  $x=S\frac{\Gamma\left(\frac{1}{d}+n\right)}{\Gamma(n)}$.
The closed result in (\ref{calculation_GFF_l-th-thermodynamic_limit}), valid for $n\ll N$, is explicitly given in dimensions $d=1,2$ in Table \ref{tab:gff_components}, and represented in Fig. \ref{fig:gff_decomposition} for various choices of the neighborhood degree $n$.

\begin{table}[t]
\centering
\renewcommand{\arraystretch}{1.5} % Adjust row spacing
\setlength{\tabcolsep}{12pt} % Increase spacing between columns
\begin{tabular}{@{}l@{\extracolsep{\fill}}c@{}} % Stretch the table width
\toprule
 & \({\rm SFF}^{(n)}(k)\) \\
\midrule
$d=1$ & \(\hspace{20pt} \frac{1}{N}\left(k^{2}\rho^2+1\right)^{-\frac{n}{2}} \cos\left[n\arctan(k\rho)\right] \) \\[5pt]
$d=2$ & \(\hspace{20pt} \frac{1}{N} \, _1F_1\left(n,\frac{1}{2};-\frac{k^{2}\rho^2}{4}\right)\) \\[5pt]
\bottomrule
\end{tabular}
\caption{The $n$-th component of the spatial form factor,  
${\rm SFF}^{(n)}(k)$, for a BPP of $N$ points on a line ($d=1$), and on a disk ($d=2$). The factor $\rho$ equals the ratio $R/N^{\frac{1}{d}}$. The expressions are valid in the large $N$ limit.}
\label{tab:gff_components}
\end{table}

As a remark, we note that the one-dimensional ${\rm SFF}$ and its decomposition can be equivalently mapped onto the spectral form factor $K(t)$ of an integrable system, where the energy level spacings follow Poissonian statistics \cite{shir2023range}. The difference between the components of the $\mathrm{SFF}$ and $K(t)$ is solely in the pre-factor, which arises from the different definitions of neighbors previously discussed.
The correspondence extends to certain components of $K(t)$ for random matrices from Gaussian ensembles (GOE, GUE, GSE) and the components of $\mathrm{SFF}$ for a PPP in 2D. This is due to the equivalence between the spacing distributions of the $n$-th nearest neighboring eigenvalues in Gaussian ensembles and those in a two-dimensional PPP \cite{Sakhr2006}.

For completeness, we also report the expression for ${\rm SFF}^{(n)}(k)$ in dimensions $d=3$, which reads
\begin{align}
&\mathrm{SFF}^{(n)}_{\text{$d=3$}}(k)\nonumber\\
&= \frac{1}{N(n-1)!} \left[\Gamma (n) \, {}_2F_5\left(\frac{n}{2}+\frac{1}{2},\frac{n}{2};\frac{1}{6},\frac{1}{3},\frac{1}{2},\frac{2}{3},\frac{5}{6};-\frac{\rho^6 k^6}{11664}\right) \right. \nonumber \\
& \left. +\frac{\rho^4 k^4}{24} \Gamma \left(n+\frac{4}{3}\right) \, {}_2F_5\left(\frac{n}{2}+\frac{2}{3},\frac{n}{2}+\frac{7}{6};\frac{5}{6},\frac{7}{6},\frac{4}{3},\frac{3}{2},\frac{5}{3};-\frac{\rho^6 k^6}{11664}\right) \right. \nonumber \\
& \left. -\frac{\rho^2 k^2}{2} \Gamma \left(n+\frac{2}{3}\right) \, {}_2F_5\left(\frac{n}{2}+\frac{1}{3},\frac{n}{2}+\frac{5}{6};\frac{1}{2},\frac{2}{3},\frac{5}{6},\frac{7}{6},\frac{4}{3};-\frac{\rho^6 k^6}{11664}\right) \right].
\end{align}

\subsection{The infinite-dimensional limit}
We close this section by reporting a simple corollary obtained by considering the limit $d\rightarrow \infty$ of the SFF expression given in Eq. (\ref{finalsimplifiedSFFdBall}).
In particular, let us compute
\begin{align}
     &\mathcal{I}_{\infty}(k):=\lim_{d\to\infty}\mathcal{I}_{d}(k)\nonumber\\
     &=\lim_{d\to\infty} {}_2F_3\left(\frac{d+1}{2},\frac{d}{2};\frac{1}{2},\frac{d+2}{2},d+1;-k^2R^{2}\right). 
\end{align}
By using the series representation of the hypergeometric function \cite{NIST:DLMF}, the limit takes the form 
\begin{equation}\label{LimitFregularized}
\mathcal{I}_{\infty}(k)=\lim_{d\rightarrow \infty} 
\sum_{n=0}^{\infty}\frac{\left(\frac{d+1}{2}\right)_{n} \left( \frac{d}{2}\right)_{n}}{\left( \frac{1}{2}\right)_{n} \left(\frac{d+2}{2}\right)_{n} \left( d+1\right)_{n}}\frac{(-k^{2}R^2)^n}{n!},
\end{equation}
where $(x)_{n}$ is the Pochhammer's symbol, defined as \cite{Gradshteyn96}
\begin{equation}
    (x)_{n}=x(x+1)\cdots(x+n-1).
\end{equation}
In Eq. (\ref{LimitFregularized}), it is possible to interchange the limit with the infinite summation by virtue of Tannery's theorem \cite{Tannery_theorem}.
To apply the theorem, it is sufficient to prove that 
\begin{equation}
    \left|\frac{\left(\frac{d+1}{2}\right)_{n} \left( \frac{d}{2}\right)_{n}}{\left( \frac{1}{2}\right)_{n} \left(\frac{d+2}{2}\right)_{n} \left( d+1\right)_{n}}\frac{(-k^{2}R^2)^n}{n!}\right|\leq M_{n} \quad \text{and} \quad \sum_{n=0}^{\infty} M_{n}<\infty,
\end{equation}
with $M$ independent of $d$.
The bound is achieved by
\begin{equation}
    \left|\frac{\left(\frac{d+1}{2}\right)_{n} \left( \frac{d}{2}\right)_{n}}{\left( \frac{1}{2}\right)_{n} \left(\frac{d+2}{2}\right)_{n} \left( d+1\right)_{n}}\frac{(-k^{2}R^2)^n}{n!}\right| \leq \frac{2(k^{2}R^{2})^{n}}{n!}
\end{equation}
and 
\begin{equation}
    \sum_{n=0}^{\infty}\frac{2(k^{2}R^{2})^{n}}{n!}=2e^{k^{2}R^{2}}<\infty.
\end{equation}
Therefore, given that
\begin{equation}\label{limitPochammerfunctions}
    \lim_{d\rightarrow \infty} \frac{\left(\frac{d+1}{2}\right)_{n} \left( \frac{d}{2}\right)_{n}}{\left( \frac{1}{2}\right)_{n} \left(\frac{d+2}{2}\right)_{n} \left( d+1\right)_{n}}=\frac{2^{-n}}{(\frac{1}{2})_{n}},
\end{equation}
bringing the limit inside the sum in Eq. (\ref{LimitFregularized}), we obtain
\begin{equation}
\mathcal{I}_{\infty}(k)=\sum_{n=0}^{\infty}\frac{2^{-n}}{(\frac{1}{2})_{n}} \frac{(-k^{2}R^2)^n}{n!}=\cos(\sqrt{2}kR).
\end{equation}
Recalling the definition of $\mathcal{I}_{d}$ from Eq. (\ref{Id_2}), one finds the following equality, valid in the $d\rightarrow \infty$ limit:
\begin{align}\label{eqcosexpansions}
&\langle \cos(ks)\rangle=\cos(\sqrt{2}kR)\nonumber\\
&\implies \sum_{n=0}^{\infty} \frac{(-1)^n k^{2n}}{(2n)!} \langle s^{2n} \rangle=\sum_{n=0}^{\infty} \frac{(-1)^n k^{2n} (\sqrt{2}R)^{2n}}{(2n!)}.
\end{align}
By repeatedly taking derivatives with respect to $k$ on both sides of Eq. (\ref{eqcosexpansions}) and evaluating it at $k=0$, we see that
\begin{equation}\label{even_moments_distance_d_infty}
\langle s^{2n}\rangle= (\sqrt{2} R)^{2n}\quad \text{,} \quad n \in \mathbb{N}.
\end{equation}
Equation (\ref{even_moments_distance_d_infty}) provides the limit of all even moments of the distance between two random, uniformly distributed, and independent points sampled within a \(d\)-ball as \(d\) approaches infinity. A similar derivation, using the sine function instead of the cosine to define the \(\mathrm{SFF}\), leads to the odd moments of the distance:
\begin{equation}
   \langle s^{2n+1}\rangle =(\sqrt{2}R)^{2n+1} \quad \text{,} \quad n \in \mathbb{N}.
\end{equation}
In this case, it can be shown that the moments completely determine the probability distribution given that (see \cite{Feller1991}, p. 514), 
\begin{equation}
   \lim_{n\rightarrow\infty} {\rm sup}\,\frac{1}{n}\langle s^{n}\rangle^{\frac{1}{n}}=0<\infty.
\end{equation}
The distance distribution when $d\rightarrow \infty$ is the Dirac delta function 
\begin{equation}
   P_{\infty}(s) =\delta(s-\sqrt{2}R),
\end{equation}
which indicates a concentration of measure, with the suppression of statistical fluctuations.

\section{Spatial Form Factor of a Coulomb Gas}\label{sec:SFF_Coulomb}
Up to this point, we have primarily focused on the SFF of Poissonian point processes, where points are sampled independently of each other.
This independence can be interpreted as the absence of interaction between the points, making this scenario a model for systems of non-interacting particles. Consequently, the probability density of any given point configuration is separable into the product of the individual probability densities corresponding to each point.
A natural generalization is to extend the analysis to point processes where interactions are present. A notable example is given by a system of points subject to Coulomb interaction \cite{serfaty2018}.
In this section, we investigate the ${\rm SFF}$ in such a setting.

Let us consider, as before, a set of \( N \) points with positions denoted by \((\mathbf{r}_{1}, \dots, \mathbf{r}_{N})\), defined over a generic domain \(\mathcal{D}\). In addition, each point carries an electric charge \( q = \pm 1 \). The Coulomb interaction between any pair of points at positions \(\mathbf{r}_{i}\) and \(\mathbf{r}_{j}\), with respective charges \( q_{i} \) and \( q_{j} \), is given by the pair potential:
\begin{equation}\label{Coulomb_V}
V(\mathbf{r}_{i}, \mathbf{r}_{j}) =
\begin{cases}
    -q_{i}q_{j}|\mathbf{r}_{i} - \mathbf{r}_{j}| & \quad d = 1, \\
    -q_{i}q_{j}\log(|\mathbf{r}_{i} - \mathbf{r}_{j}|) & \quad d = 2, \\
    q_{i}q_{j}\frac{1}{|\mathbf{r}_{i} - \mathbf{r}_{j}|^{d-2}} & \quad d > 2.
\end{cases}
\end{equation}

Due to the interaction described by (\ref{Coulomb_V}), if all the points in the system have the same charge, they will repel each other and eventually escape to infinity. To prevent this, an additional confining external potential \( V_{\text{ext}} \) is applied. A typical choice for \( V_{\text{ext}} \) is a harmonic trap. This choice also arises naturally in the $d=2$ case in the discussion of log gases in relation to random matrix theory \cite{Forrester2010}. It also arises in $d=1$ in the characterization of the ground-state density of many-body quantum models such as the long-range Lieb-Liniger Hamiltonian \cite{delcampo20conf,Beau20} and related diffusion models \cite{LeDoussal22}.
In our study, we employ a box trap potential to preserve the homogeneity of the system.

The probability density of a certain $N$-point configuration is known from the canonical ensemble in statistical mechanics and given by the Gibbs measure \cite{kardar2007statistical}
\begin{equation}
    P(\mathbf{r}_{1},\dots,\mathbf{r}_{N})=\frac{1}{\mathcal{Z}}e^{-\frac{\beta}{2}\sum_{i\neq j=1}^{N}V(\mathbf{r}_{i},\mathbf{r}_{j})},
\end{equation}
where $\beta$ plays the role of the inverse temperature of the system, and $\mathcal{Z}$ is the classical canonical partition function
\begin{equation}\label{Z}
    \mathcal{Z}=\int_{\mathcal{D}^N}  e^{-\frac{\beta}{2}\sum_{i\neq j=1}^{N}V(\mathbf{r}_{i},\mathbf{r}_{j})} \,d^{N}\mathbf{r}.
\end{equation}
The spatial form factor for a system of points with interaction (\ref{Coulomb_V}) is given by
\begin{align}\label{GFF_coulomb}
&\mathrm{SFF}(k) = \frac{1}{N^2} \left< \sum_{i,j=1}^{N} \cos(|\mathbf{r}_{i}-\mathbf{r}_{j}|k) \right> \nonumber\\
&= \frac{1}{N^2}\left(N+\frac{1}{\mathcal{Z}}\int_{\mathcal{D}^{N}} \sum_{i\neq j=1}^{N}\cos(|\mathbf{r}_{i}-\mathbf{r}_{j}|k) e^{-\frac{\beta}{2} \sum_{n\neq m=1}^{N} V(\mathbf{r}_{n},\mathbf{r}_{m})} \,d^{N}\mathbf{r}\right).
\end{align}

Obtaining a closed form of (\ref{GFF_coulomb}) is generally challenging.
Therefore, in the following subsection, we concentrate on the one-dimensional case and assume the regime $\beta \rightarrow 0$ (high-temperature limit) to determine the first-order correction to the ${\rm SFF}$.
In the subsequent subsection, we investigate the $\mathrm{SFF}$ for the Coulomb gas in higher dimensions using Monte Carlo simulations.

\subsection{High temperature expansion of the spatial form factor of a 1$d$ Coulomb gas}\label{perturbative_SFF_1d_coulomb}
Let us consider a 1$d$ Coulomb gas confined on the segment $[-R,R]$. For now, we assume that the system consists of $N$ particles, with $N_{+}$ positively charged and $N_-=N-N_+$ negatively charged, all with equal magnitude $q=|q_n|$.
The ${\rm SFF}$ of the system, according to Eq. (\ref{GFF_coulomb}), is given by
\begin{equation}\label{GFF_1d_coulomb}
     {\rm SFF}(k)=\frac{1}{N^{2}}\left(N+\frac{1}{\mathcal{Z}}\mathcal{C}(k,\beta)\right),
\end{equation}
where, to simplify the expression, we have defined $\mathcal{C}(k, \beta)$ as 
\begin{equation}\label{C_k_beta_def}
    \mathcal{C}(k, \beta)=\int_{[-R,R]^N} \sum_{i\neq j}\cos(|x_{i}-x_{j}|k) e^{\frac{\beta}{2} (\sum_{n\neq m} q_{n}q_{m} |x_{n}-x_{m}|)}\,d^{N}x.
\end{equation}
The indices $(i, j)$ and $(n, m)$ range from $1$ to $N$, but this has been omitted here and in what follows for brevity.

\begin{figure}[t]
    \centering
\includegraphics[width=0.8\linewidth]{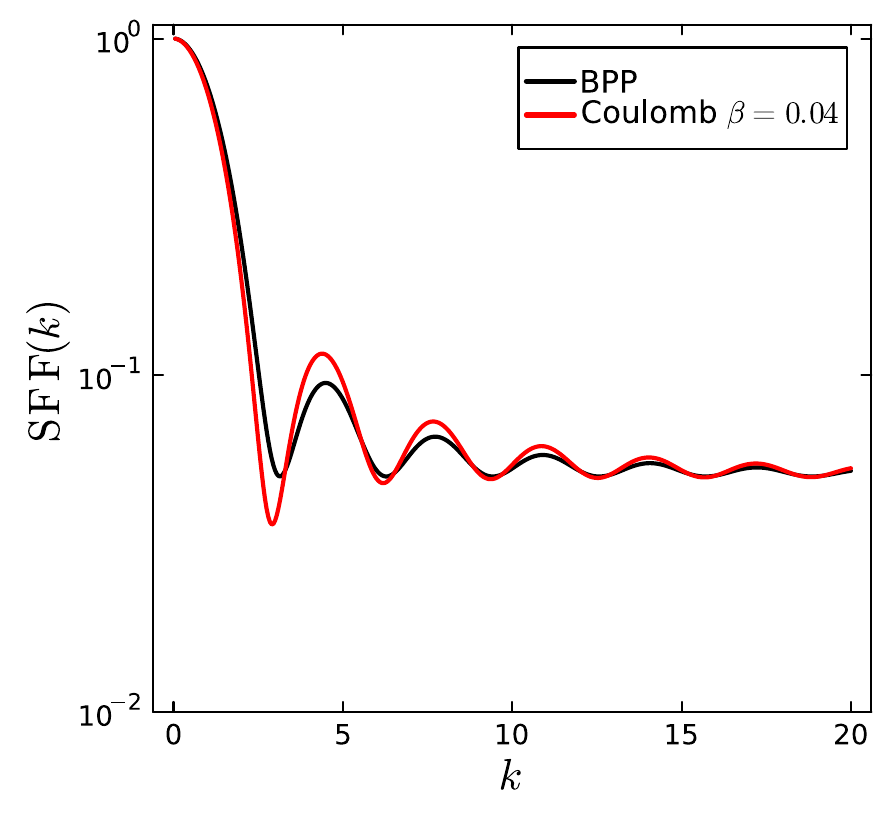}
    \caption{Comparison of the $\rm SFF$ for a BPP, corresponding to \( \beta=0 \), and a Coulomb gas with \( \beta=0.04 \) in 1D. The system consists of \( N = 20 \) particles confined to the segment \( [-1, 1] \), with each particle carrying a charge $q=+1$.
    The red curve represents the analytic expression for the Coulomb gas, given by Eq. (\ref{final_result_GFF_Coulomb_beta_linear}), while the black curve corresponds to the analytic $\rm{SFF}$ of a BPP, as derived in Sec. \ref{GFFPPP}. A logarithmic scale is used on the $y$-axis to highlight the differences between the two curves.
  }
    \label{fig:gff_coulomb_low_beta}
\end{figure}
We now focus on the regime $\beta \rightarrow 0$ and compute the linear correction in $\beta$ to the $\mathrm{SFF}$. The detailed perturbative expansion is presented in Appendix~\ref{appendA}. 

The resulting expression for the $\mathrm{SFF}$, valid up to linear order in $\beta$, is

\begin{align}\label{final_result_GFF_Coulomb_beta_linear}
    &{\rm SFF}(k)=\frac{1}{N^2}\left(N+\frac{1}{\mathcal{Z}}\mathcal{C}(k,\beta)\right)\nonumber\\
    &=\mathrm{SFF}_{\scaleto{(\rm{BPP)}\mathstrut}{5pt}}(k)\nonumber\\
    &+\frac{\beta}{N^{2}(2R)^{N}}\left[\frac{a(k)+b(k)+c(k)}{2}\right. \nonumber\\
    &\qquad \quad \left.-\frac{4}{3}N(N-1)(2R)^{(N-1)}q^{2}(P_{+}-P_{-})\frac{\sin^{2}(kR)}{k^{2}}\right],
\end{align}
where
\begin{equation}
    P_{+}=\binom{N_+}{2}+\binom{N_-}{2} \quad \text{,} \quad P_{-}=N_{+}N_{-}.
\end{equation}
The explicit expressions for the functions $a(k)$, $b(k)$, and $c(k)$ are somewhat lengthy and provided in Appendix~\ref{appendA}.

In Fig. \ref{fig:gff_coulomb_low_beta}, we consider the case of a Coulomb gas with all the particles carrying the same charge $q=+1$, at inverse temperature $\beta=0.04$.
In particular, we compare the plot of the perturbative expression for the $\mathrm{SFF}$ given in Eq. (\ref{final_result_GFF_Coulomb_beta_linear}) with that of the corresponding BPP.

\begin{figure*}[t]
    \centering
\includegraphics[width=0.9\linewidth]{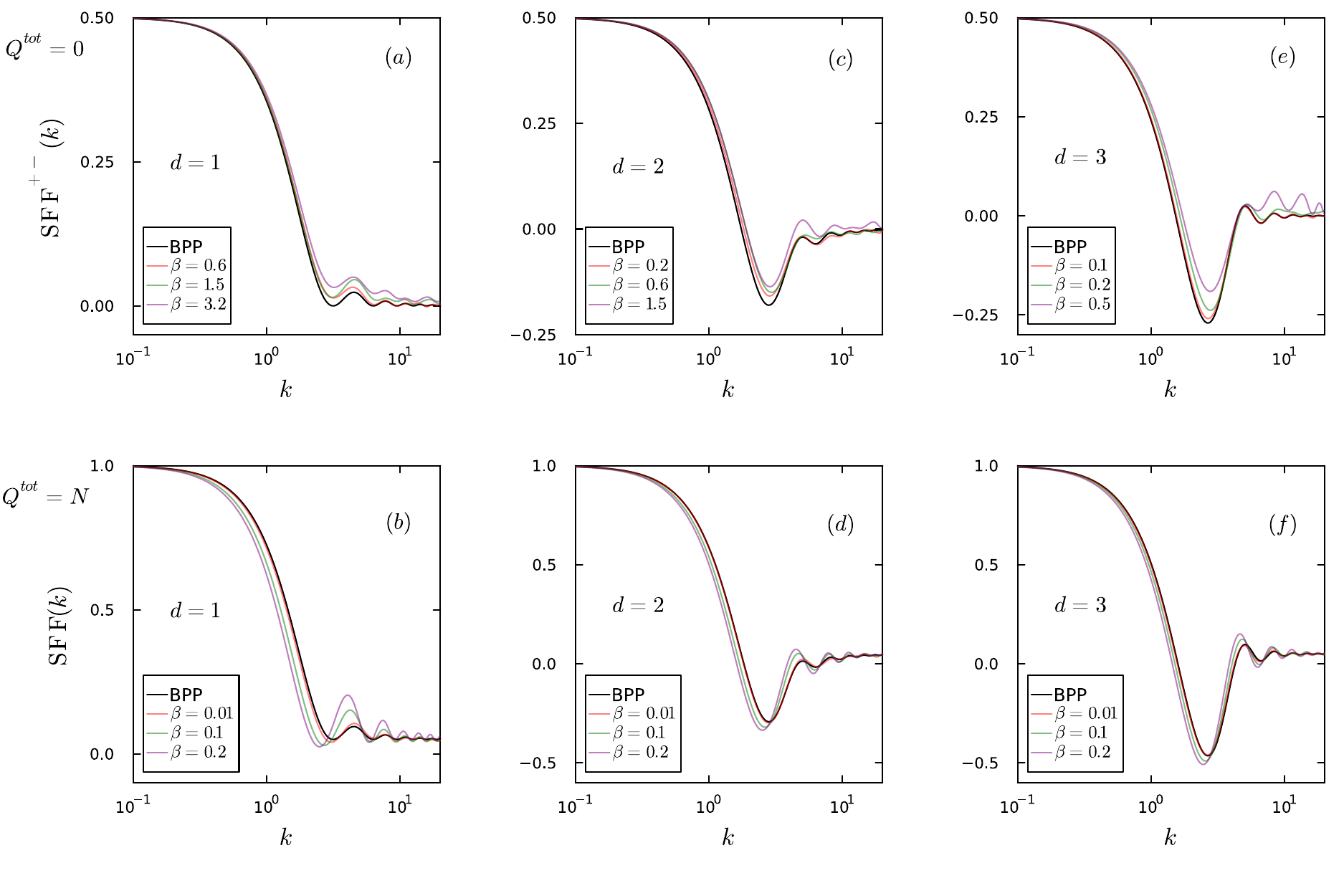}
    \caption{Spatial form factor of a system of \(N=20\) points with Coulomb interaction for different values of the inverse temperature \(\beta\) in dimensions \(d = 1, 2, 3\). The 1D system is represented by a segment of length \(2R=2\), while the 2D and 3D systems are represented by a disk and a sphere of unit radius, respectively. Panels (a), (c), and (e) correspond to a neutral Coulomb gas, depicting the component \(\mathrm{SFF}^{+-}\) of the spatial form factor, which considers only the distances between opposite charges as per the decomposition in Eq. (\ref{GFF_Coulomb_decomposition}). Panels (b), (d), and (f) correspond to a gas of all positive charges, showing the corresponding full \(\mathrm{SFF}\). In each panel, the black line represents the analytic expression for \(\mathrm{SFF}\) of the corresponding BPP, while the colored lines have been determined by a Monte Carlo simulation, averaging over $5\times 10^{3}$ configurations.
}
    \label{fig:gff_coulomb}
\end{figure*}

Equation (\ref{final_result_GFF_Coulomb_beta_linear}) shows that when $\beta=0$, we recover the BPP case, a feature shared by any other interacting point process associated with a bounded potential, i.e., 
\begin{align}
    \lim_{\beta\rightarrow0 }{\rm SFF}(k)=
    \mathrm{SFF}_{\scaleto{(\rm{BPP)}\mathstrut}{5pt}}(k).
\end{align}
The last line of Eq. (\ref{final_result_GFF_Coulomb_beta_linear}) gives the first order correction in $\beta$ for the Coulomb gas and leads to an increase of the amplitude of the oscillations of the SFF as a function of $k$.

\subsection{Spatial form factor of a Coulomb gas in higher dimensions}

In this section, we extend the analytical result of the previous section using a numerical approach.

Since the charge labels the points, it is useful to decompose the $\mathrm{SFF}$ as 
\begin{align}\label{GFF_Coulomb_decomposition}
\mathrm{SFF}(k) &= \frac{1}{N^2} \left< \sum_{i=1}^{N_{+}} \sum_{j=1}^{N_{+}} \cos(|\mathbf{r}_{i}-\mathbf{r}_{j}|k)  + \sum_{i=1}^{N_{-}} \sum_{j=1}^{N_{-}} \cos(|\mathbf{r}_{i}-\mathbf{r}_{j}|k) \right. \nonumber\\
&\quad \left. + 2\sum_{i=1}^{N_{+}} \sum_{j=1}^{N_{-}} \cos(|\mathbf{r}_{i}-\mathbf{r}_{j}|k) \right> \nonumber\\
&= \mathrm{SFF}^{++}(k) + \mathrm{SFF}^{--}(k) + \mathrm{SFF}^{+-}(k),
\end{align}
where the first two terms, $\mathrm{SFF}^{++}$ and $\mathrm{SFF}^{--}$, account for points with the same charge, positive and negative respectively, whereas $\mathrm{SFF}^{+-}$ is computed by considering only the spacing between pairs of points with opposite charges.
Figure \ref{fig:gff_coulomb} shows the $\mathrm{SFF}$ numerically computed for the Coulomb gas. Specifically, Panels (a), (c), and (e) display the component $\mathrm{SFF}^{+-}$ for a neutral Coulomb gas in different dimensions. Panels (b), (d), and (f) depict the $\mathrm{SFF}$ of a Coulomb gas where all the charges are positive, in different dimensions.
As expected, when the value of the inverse temperature $\beta$ tends to zero, the interaction vanishes, and we recover the BPP case. This can, in fact, be seen as the limiting case of a Coulomb gas at infinite temperature, where the thermal agitation is so intense that the interaction between points does not play a role. 

As discussed in Sec. \ref{SecCharGFF}, the $\mathrm{SFF}$ for a BPP in $d > 1$ dimensions exhibits a feature analogous to the correlation hole observed in the spectral form factor of a quantum system, which arises due to the presence of energy level repulsion.
Similarly, this phenomenon in the $\mathrm{SFF}$ can be attributed to the behavior of the nearest neighbor spacing distribution $P^{\text{\tiny NN}}_{d}(S)$, given in Eq. (\ref{eq:NN_spacing_PPP_d_ball}), which vanishes for small $S$, giving rise to an apparent point repulsion. 

In a Coulomb gas with $\beta>0$, as demonstrated in Fig. \ref{fig:gff_coulomb}, this effect can be either enhanced or diminished depending on the distribution of particle charge signs. Specifically, for a Coulomb gas with equal charges of the same sign, the leading correction in $\beta$ leads to a faster decay from the unit value, reducing the value of the dip and enhancing the amplitude of the oscillations following it. This can be seen as a consequence of the repulsive interaction. On the other hand, in a neutral Coulomb gas, the component $\mathrm{SFF}^{+-}$ exhibits a less pronounced dip. This results from the attraction between particles with opposite charges, which counteracts the point repulsion.

\section{Vortex Formation during Bose-Einstein Condensation} \label{Sec:Vortex_Formation}

\begin{figure*}
    \centering
    \includegraphics[width=0.75\linewidth]{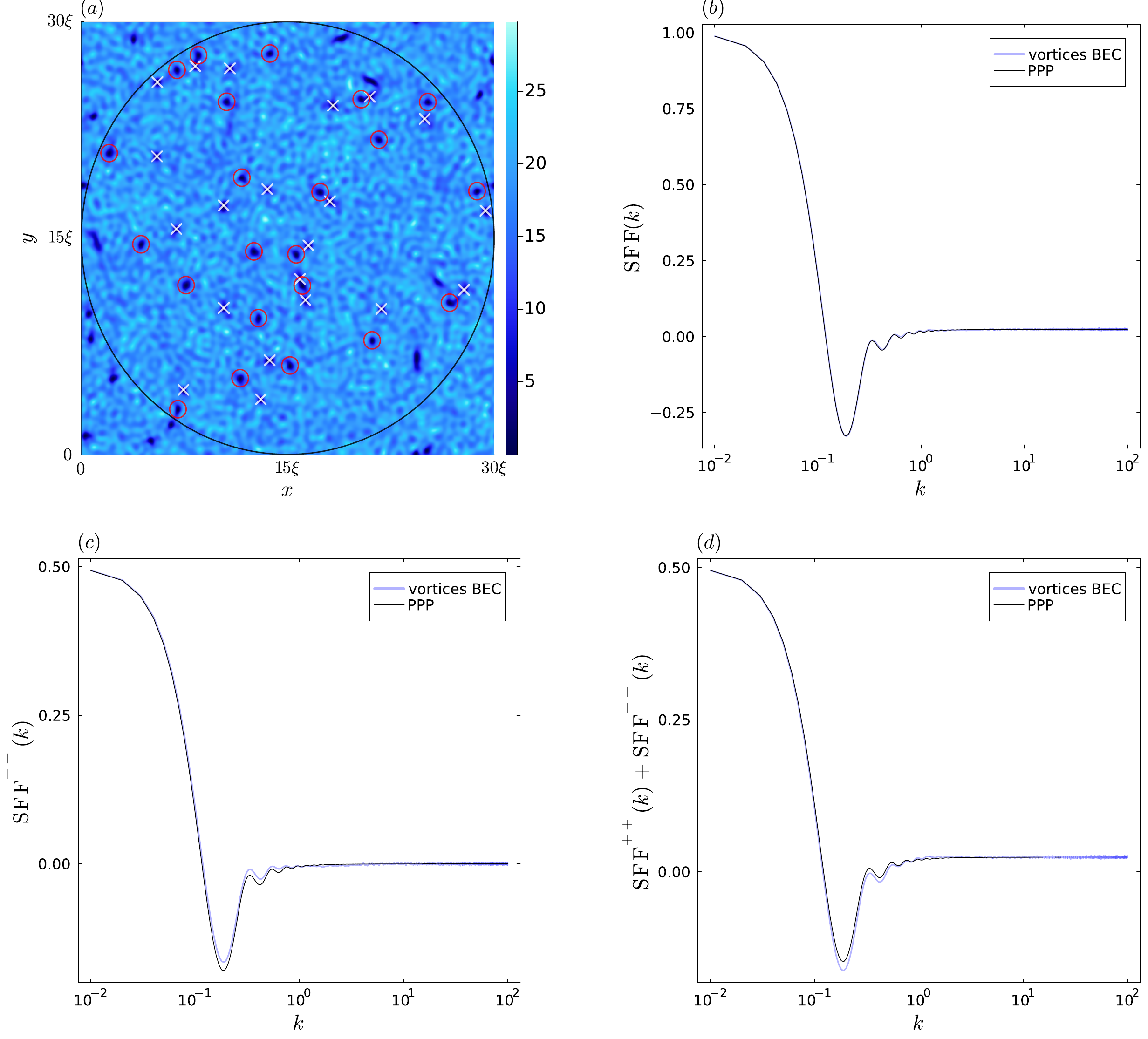}
    \caption{Panel (a) depicts an example of a typical realization of the BEC density $|\Psi(x,y)|^{2}$ after a fast quench ($\tau_{Q}=20$) described by the Stochastic Projected Gross-Pitaeveskii equation. Vortices appear as zeroes of the BEC density profile associated with quantized circulation and $\pm 1$ charge, respectively marked with a red cross and a white circle. The coordinates of the vortices constitute a random point pattern encoding the stochastic geometry of spontaneous symmetry breaking.
    Panels (b), (c) and (d) show the characterization of the spatial statistics of spontaneous vortex patterns within a circular region of the system domain using the SFF. 
    The numerical results for the SFF of vortex patterns, averaged over 361 realizations, are compared with the analytical expression of the  SFF for a corresponding PPP as given in Table \ref{tab:I_d_1_2_3}. The agreement is excellent when the charge of the vortices is ignored. Small deviations are noticeable when conditioning the SFF on the vortex charge according to Eq. (\ref{GFF_Coulomb_decomposition}).}
    \label{fig:density_BEC_and_SFF_vortices_vs_PPP}
\end{figure*}

As a further example of the utility of the SFF, we consider the characterization of the spatial patterns of vortices spontaneously formed during a thermal quench cooling an ultracold gas below the temperature for Bose-Einstein condensation. The breaking of $U(1)$ symmetry in finite time leads to the proliferation of vortices, that can be imaged in ultracold atom experiments \cite{Weiler08,Navon2015,Ko2019,Goo2021,Goo2022,lee2023observation}. As the dynamics is stochastic, the coordinates of the vortices in each realization constitute a random point pattern. That such patterns are accurately described as a PPP with density dictated by the Kibble-Zurek mechanism has been recently established \cite{delcampo22,thudiyangal2024}.

In what follows, we illustrate the use of the SFF to reveal the stochastic geometry of such patterns.
The time evolution of the condensate wave function $\Psi(\mathbf{r})$ is governed by the Stochastic Projected Gross-Pitaevskii equation \cite{Gardiner_2003, review_Gardiner, blakie2008, bradley2015}, which in dimensionless units reads 
\begin{equation}
d{\Psi} = \mathcal{P}_{C} \left[ -(i + \gamma) \left( H_{sp} + g |\Psi|^{2} -\epsilon(t) \right) \Psi \, dt + d \eta \right].
\end{equation}
Here, $H_{sp}$ denotes the single particle Hamiltonian, and $d\eta$ a complex Gaussian white noise with correlator
\begin{equation}
    \langle d\eta (\mathbf{r},t) d\eta^{*} (\mathbf{r}^{\prime},t)\rangle=2T\gamma \delta(\mathbf{r}-\mathbf{r}^{\prime})dt,
\end{equation}
where $\gamma$ is the dissipation rate and $T$ the temperature of the gas.
In our simulations we considered a homogeneous $L\times L$ system with periodic boundary conditions. 
The units of length, time, energy and temperature have been respectively chosen to be $\xi=\sqrt{\frac{\hbar^2}{mgn_{0}}}$, $\frac{m\xi^2}{\hbar}$, $\frac{\hbar^2}{m\xi^2}$ and $\frac{\hbar^2}{K_{B}m\xi^2}$, with $\xi$ the condensate healing length, and $n_{0}$ the equilibrium condensate density.
The projector operator $\mathcal{P}_{C}$ restricts the dynamics to the coherent region, represented by the low-lying single particle energy modes, up to an arbitrary energy cut-off $\epsilon_{cut}$.
In the following, to simulate the spontaneous formation of vortices, we consider a linear quench of the control parameter $\epsilon(t)$:
\begin{equation}
\epsilon(t)=\mu_{i}+\frac{t}{\tau_{Q}}(\mu_{f}-\mu_{i}),
\end{equation}
where $\mu_{i}$ and $\mu_{f}$ are, respectively, the initial and final value of the chemical potential, while $\tau_{Q}$ is the quench duration.
The simulations we performed correspond to the following parameter values: $\mu_{i}=0.1$, $\mu_{f}=20$, $\epsilon_{cut}=50$. Furthermore, we set the temperature to be $T=1$, the coupling constant $g=1$, and the dissipation rate $\gamma=0.03$.

A typical realization of the BEC density profile resulting from the ramp of the chemical potential is shown in Fig.  \ref{fig:density_BEC_and_SFF_vortices_vs_PPP}. In addition to the nonequilibrium and finite-temperature modulations of the density, one can identify vortices at the locations where the density vanishes. The circulation around such zeroes of the BEC density is quantized, and the associated winding number is restricted to $\pm1$, distinguished by circles and crosses. Vortices with higher winding numbers are energetically suppressed and are also unstable against decay. The coordinates of the vortices realize a random point pattern.

We use the SFF to characterize the vortex spatial statistics. Figure \ref{fig:density_BEC_and_SFF_vortices_vs_PPP} shows the averaged SFF computed from vortex patterns generated through 361 independent trajectories of the Stochastic Projected Gross-Pitaevskii equation. This is compared with the SFF derived for a PPP with $d=2$ given in Table \ref{tab:I_d_1_2_3}.
In particular, the solid black line in Figure \ref{fig:density_BEC_and_SFF_vortices_vs_PPP} (b) corresponds to the expression
\begin{equation}
    \mathrm{SFF}(k)=\frac{1}{N}+\left(1-\frac{1}{N}\right)\mathcal{I}_{2}(k),
\end{equation}
where, to make the comparison with the SFF of the vortex positions, the value of $\frac{1}{N}$ is set equal to $
\Big\langle \frac{1}{N_v} \Big\rangle$. 
Here, $N_v$ denotes the vortex number in the circular system computed from the numerical data, and the angle brackets 
$\langle \cdot \rangle$ indicate an average over 361 independent stochastic trajectories. 
Similarly, the black solid line in Figure \ref{fig:density_BEC_and_SFF_vortices_vs_PPP} (c) corresponds to
\begin{equation}
   \mathrm{SFF}^{+-}(k)=2\left< \frac{N_{+}N_{-}}{N_{v}^2}\right> \mathcal{I}_{2}(k), 
\end{equation}
and in Figure \ref{fig:density_BEC_and_SFF_vortices_vs_PPP} (d) to
\begin{align}
    \mathrm{SFF}^{++}(k)&+\mathrm{SFF}^{--}(k)
    =\left< \frac{N_{+}}{N_{v}^2}\right>+ \left< \frac{N_{+}(N_{+}-1)}{N_{v}^{2}}\right>\mathcal{I}_{2}(k) \nonumber\\
    &+\left< \frac{N_{-}}{N_{v}^2}\right>+ \left< \frac{N_{-}(N_{-}-1)}{N_{v}^{2}}\right>\mathcal{I}_{2}(k),
\end{align}
where the value of the averaged quantities is obtained from the SPGPE data.

When the circulation of the vortices is ignored, the vortex spatial statistics is described with excellent accuracy by the PPP model, to the extent that the numerical simulations and the analytical expression are barely distinguishable. Conditioning on the charge allows for different definitions of the SFF involving vortices of different charges. The agreement with the PPP remains highly satisfactory, with only slight deviations expected from Eq. (\ref{GFF_Coulomb_decomposition}).

%\begin{figure*}
%    \centering
%    \includegraphics[width=1.0\linewidth]{fig/plot_comparison_SFF_vortices_PPP.pdf}    \caption{\textcolor{blue}{Characterization of the spatial statistics of spontaneous vortex patterns using the SFF. Bose-Einstein condensation is modeled using the Stochastic Projected Gross-Pitaevskii equation and induced by a ramp of the chemical potential. The numerical results averaged over 361 realizations are compared with the analytical expression of the SFF in a PPP given in Table \ref{tab:I_d_1_2_3}. The agreement is excellent when the charge of the vortices is ignored. Small deviations are noticeable when conditioning the SFF on the vortex charge according to Eq. (\ref{GFF_Coulomb_decomposition}).}}
%    \label{fig:SFFBEC}
%\end{figure*}

\section{Conclusions and Discussion}\label{SecSummary}

In this work, we have introduced the spatial form factor for the characterization of spatial point patterns and detailed the analysis for random patterns. We have established two fundamental results. First, the SFF is related to the even Fourier transform of the probability distribution for the distance between any pair of points, which is familiar from the ball line picking problem in stochastic geometry \cite{mathai1999introduction,Chiu2013stochastic}. Second, we have established the relation between the SFF and the set of $n$-order spacing distributions. We have focused on the explicit computation of the SFF for homogeneous Poisson point processes in arbitrary dimensions, for which we have provided explicit expressions in terms of well-known functions.

The line-picking problem has been studied and finds application in many other enclosed regions in $\mathbb{R}^d$, such as convex polytopes, suggesting further extensions of our work to such geometries \cite{mathai1999introduction,Chiu2013stochastic}.

As a case study, we have shown how the SFF can be used to characterize vortex formation during Bose-Einstein condensation. Via the  Kibble-Zurek mechanism \cite{delcampo2014}, the critical dynamics leads to the spontaneous formation of random spatial patterns of vortices whose spatial statistics is described by a PPP \cite{delcampo22,thudiyangal2024}, as we have verified by analyzing the corresponding SFFs.

%With an eye on applications in nonequilibrium statistical mechanics and condensed matter physics, we have applied the SFF   to the characterization of random spatial patterns of point-like topological defects such as vortices, generated via the Kibble-Zurek mechanism \cite{delcampo2014}, for which the accuracy of the Poisson point process has been established \cite{delcampo22,thudiyangal2024}.} 

A natural extension of our work involves the analytical study of the role of interactions that can be either contact (e.g.,  soft or hard-core interactions between particles of finite radius) or with finite range. For instance, vortices on a plane are known to interact as a 2D Coulomb gas, motivating the evaluation of the SFF in such a setting, which we have discussed using a numerical approach.
 Another extension motivated by the possibility of probing vortex spatial statistics in experiments with ultracold gases \cite{Weiler08,KimShin22} is the inclusion of an external potential, which can make the spatial patterns inhomogeneous, according to the Inhomogeneous Kibble-Zurek mechanism \cite{Zurek09,DRP11, DKZ13,KimShin22}.
Beyond these applications, the SFF should prove useful in analyzing and characterizing point processes in general, including the case of regular and quasiperiodic patterns, as proposed in Sec. \ref{SecGFFdef}. 

Finally, we note that the SFF can be generalized as
\begin{equation}
{\rm SFF}(k)=\left\langle \frac{1}{|X|^2}\sum_{x,y\in X}\cos[d(x,y)k]\right\rangle     
\end{equation}
to an arbitrary metric space $(M,d)$ \cite{Burago01} 
 with a distance $d(x,y)$ between points $x$ and $y$ of a set $X\subseteq M$ with finite cardinality $|X|<\infty$. Such generalization applies to other fields, such as information theory \cite{Cover12}, making it possible to analyze, e.g., sets of strings using the Hamming distance. In the same vein, it could be applied in information geometry to characterize the distance distribution between a set of probability distributions \cite{Amari16} and in quantum information theory for exploring the distance distribution of quantum states (or even quantum operations) in a given set \cite{Hayashi06}.

\acknowledgements
The authors are indebted to Pablo Mart\'inez-Azcona, Ruth Shir, Andr\'{a}s Grabarits, and Seong-Ho Shinn for insightful discussions and comments on the manuscript. 
This research was funded by the Luxembourg National Research Fund (FNR), grant reference 17132060. 
%For the purpose of open access, the author has applied a Creative Commons Attribution 4.0 International (CC BY 4.0) license to any Author Accepted Manuscript version arising from this submission.

\appendix

\section{Small $\beta$ expansion of the spatial form factor of a 1$d$ Coulomb gas}\label{appendA}

In this appendix, we perform the small $\beta$ perturbative expansion of the $\mathrm{SFF}$ of the 1$d$ Coulomb gas, as presented in Section~\ref{perturbative_SFF_1d_coulomb}. Specifically, we determine the linear correction in $\beta$ to the $\mathrm{SFF}$.

We proceed by expanding the partition function $\mathcal{Z}$, defined in Eq. (\ref{Z}), as 
\begin{align}\label{Calculation_Z}
    &\mathcal{Z}=\int_{[-R,R]^N}\left( 1+\frac{\beta}{2}\sum_{i\neq j} q_{i} q_{j} |x_{i}-x_{j}| \right) d^{N}x+O(\beta^{2})\nonumber\\
    &=(2R)^{N}+\beta\left(P_{+}-P_{-}\right)q^{2}(2R)^{N-2}\nonumber\\
    &\qquad \qquad \times\int_{[-R,R]^2}  |x_{1}-x_{2}| dx_{1} dx_{2}+O(\beta^{2}) \nonumber\\
    &=(2R)^{N}+\beta (P_{+}-P_{-}) q^{2}(2R)^{(N-2)} \frac{8}{3}R^3 +O(\beta^{2}),
\end{align}
where $P_{+}$ denotes the number of particle pairs with the same charge sign, and $P_{-}$ the number of pairs with opposite charges. Specifically,
\begin{equation}
    P_{+}=\binom{N_+}{2}+\binom{N_-}{2} \quad \text{,} \quad P_{-}=N_{+}N_{-}.
\end{equation}
Consequently, $\mathcal{Z}^{-1}$ is given by
\begin{equation}\label{Z_inverse}
    \frac{1}{\mathcal{Z}}=\frac{1}{(2R)^N}\left(1-\frac{2}{3}q^{2}R(P_{+}-P_{-}) \beta\right)+O(\beta^2).
\end{equation}

Let us now carry out the same expansion for $\mathcal{C}(k,\beta)$, defined in Eq. (\ref{C_k_beta_def}):
\begin{align}\label{compute_GFF_1d_first_order_beta}
    &\mathcal{C}(k,\beta)
    =\int_{[-R,R]^N} \sum_{i\neq j}\cos(|x_{i}-x_{j}|k) \nonumber\\
    &\qquad \qquad \times\left(1 +\frac{\beta}{2}\sum_{n\neq m}q_{n}q_{m}|x_{n}-x_{m}|+O(\beta^2)\right)d^{N}x\nonumber\\
    &=\left(N(N-1)(2R)^{N-2}\int_{-R}^{R}\int_{-R}^{R} \cos(|x_{1}-x_{2}|k) \,dx_{1} dx_{2} \right)\nonumber\\
    &+\frac{\beta}{2} \left(\int_{[-R,R]^{N}}\sum_{\substack{i \neq j \\ n \neq m}}\cos(|x_{i}-x_{j}|k) q_{n}q_{m}|x_{n}-x_{m}| \,d^{N}x\right)+O(\beta^2).
\end{align}
The integral in the first parentheses of Eq. (\ref{compute_GFF_1d_first_order_beta}) can be readily computed, 
\begin{align}
    &N(N-1)(2R)^{N-2}\int_{-R}^{R}\int_{-R}^{R} \cos(|x_{1}-x_{2}|k) \,dx_{1} \,dx_{2}\nonumber\\
    &=N(N-1)(2R)^{(N-2)}\frac{4\sin^{2}(kR)}{k^2}.
\end{align}
We then focus on the term linear in $\beta$, which appears in the final line of Eq. (\ref{compute_GFF_1d_first_order_beta}). This term can be decomposed into three distinct contributions, representing different scenarios for the index pairs. Specifically, in the expression \( \cos(|x_{i} - x_{j}| k) |x_{n} - x_{m}|\), the index pairs \((i,j)\) and \((n,m)\) can either share both indices, one index, or none. To ease the notation, we define 
\begin{equation}
    f_{i,j,n,m}(k):=\int_{[-R,R]^{N}}\cos(|x_{i}-x_{j}|k)q_{n}q_{m}|x_{n}-x_{m}|\,d^{N}x.
\end{equation}
Thus, the term inside the parenthesis in the final line of Eq. (\ref{compute_GFF_1d_first_order_beta}) can be rewritten as
\begin{align}\label{separate_a_b_c}
    &\sum_{\scriptscriptstyle i \neq j, n \neq m} f_{i,j,n,m}(k)\nonumber\\
    &= \underbrace{\sum_{\substack{\scriptscriptstyle i \neq j, n \neq m \\ \scriptscriptstyle |\{i, j\} \cap \{n, m\}| = 2}} f_{i,j,n,m}(k)}_{a(k)}+ \underbrace{\sum_{\substack{ \scriptscriptstyle i \neq j, n \neq m \\ \scriptscriptstyle |\{i, j\} \cap \{n, m\}| = 1}} f_{i,j,n,m}(k)}_{b(k)} + \underbrace{ \sum_{\substack{\scriptscriptstyle i \neq j, n \neq m \\ \scriptscriptstyle |\{i, j\} \cap \{n, m\}| = 0}} f_{i,j,n,m}(k)}_{c(k)}.
\end{align}
Let us now compute $a$, $b$ and $c$ from Eq. (\ref{separate_a_b_c}). First,
\begin{align}
    &a(k)=(2R)^{(N-2)}4(P_{+}-P_{-})q^{2}\nonumber\\
    &\qquad \quad \times\int_{[-R,R]^2}\cos(|x_{1}-x_{2}|k)|x_{1}-x_{2}|\,dx_{1} \,dx_{2}\nonumber\\
    &=(2R)^{(N-2)}4(P_{+}-P_{-})q^{2}\frac{8\cos^{2}(kR)\left[-kR+\tan(kR)\right]}{k^3}.
\end{align}
The prefactor \( 4(P_{+} - P_{-}) \) arises for the following reason. There are $2P_{+}$ terms where $(n,m)$ are chosen such that $q_{n}q_{m}=q^2$, and $2P_{-}$ where $q_{n}q_{m}=-q^2$. Since we are considering the case where $(i,j)$ and $(n,m)$ share both index values, selecting $n$ and $m$ also determines $i$ and $j$. An additional factor of 2 accounts for the possible swap of values between $i$ and $j$.

Similarly, we can compute \(b(k)\):
\begin{align}
    &b(k)=8(P_{+}-P_{-})(N-2)q^{2}(2R)^{N-3}\nonumber\\
    &\qquad\times\int_{[-R,R]^3}\cos(|x_{1}-x_{2}|k)|x_{1}-x_{3}| \,dx_{1} \,dx_{2} \,dx_{3}\nonumber\\
    &=64(N-2)(P_{+}-P_{-})q^{2}(2R)^{N-3}\nonumber\\
    &\qquad \times \sin(kR)\left(\frac{kR \cos(kR)+(k^{2}R^{2}-1)\sin(kR)}{k^4}\right).
\end{align}
Here, the factor \(8(N-2)(P_{+} - P_{-})\) is explained in a similar way as in the calculation of $a(k)$. 
Specifically, since \( (n,m) \) and \( (i,j) \) share only one index, after selecting \( (n,m) \), a factor of 2 is needed to account for the choice of the common index between \( (n,m) \) and \( (i,j) \), while $N-2$ counts the number of possibilities for the non shared index. An additional factor of 2 accounts for the possible permutations of \( (i,j) \).

Finally, \(c(k)\) is given by:
\begin{align}
    &c(k) = (N-2)(N-3)2(P_{+} - P_{-})q^{2}(2R)^{N-4}\nonumber\\
    &\times \int_{[-R,R]^4} \cos(|x_{1} - x_{2}|k) |x_{3} - x_{4}| \, dx_{1} \, dx_{2} \, dx_{3} \, dx_{4}\nonumber\\
    &= \frac{64}{3}(N-2)(N-3)(P_{+} - P_{-}) q^{2} (2R)^{N-4} R^{3} \frac{\sin^{2}(kR)}{k^{2}},
\end{align}
where the factor \((N-2)(N-3)2(P_{+} - P_{-})\) accounts for the number of possible combinations of indices $(i,j)$ and $(n,m)$, with all indices being distinct.

Plugging in Eq. (\ref{compute_GFF_1d_first_order_beta}) the results just found, we obtain the following expression for $C(k,\beta)$:
\begin{align}\label{expression_C}
\mathcal{C}(k,\beta)&= N(N-1)(2R)^{(N-2)}\frac{4\sin^{2}(kR)}{k^2}\nonumber\\
&+\frac{\beta}{2}\left[a(k)+b(k)+c(k)\right]+O(\beta^2).   
\end{align}
We now substitute into Eq. (\ref{GFF_1d_coulomb}) the expressions for $\mathcal{C}(k,\beta)$ from Eq. (\ref{expression_C}) and for the inverse partition function from Eq. (\ref{Z_inverse}), both valid at linear order in $\beta$. After performing the product and discarding the quadratic terms in $\beta$, one obtains the resulting expression for the ${\rm SFF}$ given in Eq. (\ref{final_result_GFF_Coulomb_beta_linear}).
%\begin{align}
%    &{\rm SFF}(k)=\frac{1}{N^2}\left(N+\frac{1}{\mathcal{Z}}\mathcal{C}(k,\beta)\right)\nonumber\\
%    &=\mathrm{SFF}_{\scaleto{(\rm{BPP)}\mathstrut}{5pt}}(k)\nonumber\\
%    &+\frac{\beta}{N^{2}(2R)^{N}}\left[\frac{a(k)+b(k)+c(k)}{2}\right. \nonumber\\
%    &\qquad \quad \left.-\frac{4}{3}N(N-1)(2R)^{(N-1)}q^{2}(P_{+}-P_{-})\frac{\sin^{2}(kR)}{k^{2}}\right],
%\end{align}
%as given in the main text.

\bibliography{main_v10_without_blue}
\end{document}